\newif\ifCONDITION
\CONDITIONtrue

\documentclass[review]{elsarticle}
\journal{Journal of Wind Engineering and Industrial Aerodynamics}

\usepackage{lineno,hyperref}
\usepackage{bigints}
\usepackage{ulem}
\usepackage{algorithmic}
\usepackage{setspace} 
\usepackage{amsmath,latexsym,amssymb,ascmac}
\usepackage{bm}
\usepackage{lscape}
\usepackage{mathrsfs}
\usepackage{subfigure}
\usepackage[usenames,dvipsnames]{xcolor}
\usepackage{comment}
\usepackage{algorithm}
\usepackage{graphics}
\usepackage{mathdots}
\usepackage{color}
\usepackage{soul}
\usepackage[margin=25truemm]{geometry}
\usepackage{mathtools}

\usepackage{lineno}
\newcommand*\patchAmsMathEnvironmentForLineno[1]{%
  \expandafter\let\csname old#1\expandafter\endcsname\csname #1\endcsname
  \expandafter\let\csname oldend#1\expandafter\endcsname\csname end#1\endcsname
  \renewenvironment{#1}%
     {\linenomath\csname old#1\endcsname}%
     {\csname oldend#1\endcsname\endlinenomath}}% 
\newcommand*\patchBothAmsMathEnvironmentsForLineno[1]{%
  \patchAmsMathEnvironmentForLineno{#1}%
  \patchAmsMathEnvironmentForLineno{#1*}}%
\AtBeginDocument{%
\patchBothAmsMathEnvironmentsForLineno{equation}%
\patchBothAmsMathEnvironmentsForLineno{align}%
\patchBothAmsMathEnvironmentsForLineno{flalign}%
\patchBothAmsMathEnvironmentsForLineno{alignat}%
\patchBothAmsMathEnvironmentsForLineno{gather}%
\patchBothAmsMathEnvironmentsForLineno{multline}%
}

\newcommand{\argmax}{\mathop{\rm arg~max}\limits}

\biboptions{authoryear}

\begin{document}

\begin{frontmatter}

\title{
Optimization of Sparse Sensor Placement for Estimation of Wind Direction and Surface Pressure Distribution Using Time-Averaged Pressure-Sensitive Paint Data on Automobile Model
}

 \author[label1]{Ryoma Inoba}
 \author[label1]{Kazuki Uchida}
 \author[label1]{Yuto Iwasaki}
 \author[label1]{Takayuki Nagata}
 \author[label1]{Yuta Ozawa}
 \author[label1]{Yuji Saito}
 \author[label1]{Taku Nonomura}
 \author[label1]{Keisuke Asai}

\address[label1]{Department of Aerospace Engineering, Tohoku University,\\ Sendai, Miyagi 980-8579, Japan, Email: ryoma.inoba.t1@dc.tohoku.ac.jp}

\setstretch{1.0}
\begin{abstract}
This study proposes a method for predicting the wind direction against the simple automobile model (Ahmed model) and the surface pressure distributions on it by using data-driven optimized sparse pressure sensors. Positions of sparse pressure sensor pairs on the Ahmed model were selected for estimation of the yaw angle and reconstruction of pressure distributions based on the time-averaged surface pressure distributions database of various yaw angles, whereas the symmetric sensors in the left and right sides of the model were assumed.  The surface pressure distributions were obtained by pressure-sensitive paint measurements. Three algorithms for sparse sensor selection based on the greedy algorithm were applied, and the sensor positions were optimized. The sensor positions and estimation accuracy of yaw angle and pressure distributions of three algorithms were compared and evaluated. The results show that a few optimized sensors can accurately predict the yaw angle and the pressure distributions. 
\end{abstract}
\vspace{-2cm}
\vspace{-1cm}
\end{frontmatter}
\newpage

%%%%%%%%%%%%%%%%%%%%%%%%%%%%%%%%%%%%%%%%
%%%%%%%%%%%%%%%%%%%%%%%%%%%%%%%%%%%%%%%%
%%%%%%%%%%%%%%%%%%%%%%%%%%%%%%%%%%%%%%%%
%%%%%%%%%%%%%%%%%%%%%%%%%%%%%%%%%%%%%%%%
%\pagewiselinenumbers
%\setstretch{1.6}
\setstretch{1.5}
%%%% MAIN PART
\clearpage
%\pagewiselinenumbers
%\linenumbers
%\modulolinenumbers[5]
\section{Introduction}
The automotive industry is currently undergoing a once-in-a-century transformation, and various research and technological developments are being actively conducted to address energy and environmental issues and to improve safety performance. Moreover, automated driving, electrification, and connectedness of automobiles will enable the creation of unprecedented new value. One approach to advanced automotive functionalization is the adaptive control of the vehicle and airflow around the vehicle. In a real vehicle driving environment, natural winds are present and vehicle speed varies constantly. When a vehicle runs at high speed, aerodynamic drag is the largest factor in fuel consumption \citep{hucho1993aerodynamics}, and reduction of vehicle aerodynamic drag is indispensable for improving fuel efficiency. Especially in battery-electric vehicles (BEV), which will become mainstream in the future, reducing aerodynamic drag is crucial to ensure maximum range within a given battery \citep{andwari2017review}. On the other hand, aerodynamics caused by crosswinds causes accidents due to overturning, sideslip, and rotation accidents \citep{baker1986simplified}. The action of crosswinds changes the three-dimensional structure of the flow around the vehicle and the aerodynamic characteristics \citep{krajnovic2012large,bello2016experimental, meile2016non, tunay2018experimental}. Particular attention should be paid to unsteady crosswinds, such as gusts, which generate unsteady aerodynamic forces \citep{volpe2014forces}. The appropriate control for crosswinds around a vehicle and solutions to these problems can improve driving comfortability, energy efficiency, and safety.

Adaptive vehicle control can be achieved through the steering, throttling, braking, and flow control. For flow control, active flow control that can adjust inputs appropriately is effective for adaptive control.
Active flow control is a method that uses a vehicle's electrical power to operate, which does not require any changes to the vehicle geometry. Active flow control includes various methods such as movable underbody diffuser, steady blowing, pulsed jet, steady suction, and plasma actuator, which can reduce weight and size, and thus, reduce the negative impact on vehicle performance \citep{mukut2019review}. There are many studies, both experimental and numerical, on the application of active flow control to automobiles. 
For example, \cite{mcnally2015drag} attempted to control the fluid flow by steady jets using the flat back Honda Simplified Body in numerical and experimental approaches.
\cite{bideaux2011drag} attempted to control the fluid flow by pulsed jet using the Ahmed model in an experimental approach. \cite{wassen2009road} attempted to control the fluid flow by combining suction and blowing using the Ahmed model in a numerical approach. \cite{boucinha2011drag} attempted to control the fluid flow by plasma actuators using the Ahmed model in an experimental approach. These studies have shown that active fluid control can reduce drag by 20\% \citep{mukut2019review,bideaux2011drag}.

The pressure distribution on the surface of a vehicle changes depending on the wind direction relative to the vehicle \citep{yamashita2007pressure}. This paper proposes methods for estimating the flow direction (yaw angle) and pressure distribution on a vehicle surface using pressure sensors installed around it. 
Unlike conventional airflow inclination measurement methods such as the multi-hole probe and angle of attack vane, the proposed method can reconstruct the pressure distribution and obtain more detailed information about the flow field.
Since the total weight, the space for wiring, and the installation cost of sensors are limited, it is desirable to install sparse sensors. It is necessary to optimize the position of the sensors in order to accurately estimate the yaw angle and the pressure distribution with sparse sensors. Therefore, the data-driven sparse sampling proposed by \cite{saito2021data} is applied in this paper. The data-driven sparse sampling is a technique for reconstructing full data from observation point information, and the observation point positions are optimized using training data \citep{manohar2018data,loiseau2018sparse,manohar2021optimal,saito2021determinant}. Recently, the sensor optimization methods for this purpose have been intentionally investigated by \cite{clark2018greedy,clark2020multi,clark2020sensor,yamada2021fast,yamada2021greedy,nakai2021effect,nagata2021data,nakai2022nondominated,fukami2021global,carter2021data,li2021efficient,li2021data,nagata2022randomized,nagata2022data,inoue2022data}. 
The data-driven sparse sampling method can identify and estimate the flow field at a low computational cost.  \cite{zhou2021data} succeeded in the real-time identification of inflow aerodynamic parameters, such as the flow separation situation, angle of attack, and inflow velocity. Classifying the flow field is expected to be applied to optimal feedback control \citep{deem2020adaptive,kanda2021feasibility,kanda2022proof}. In addition, \cite{kaneko2021data} attempted to reconstruct the propagation time and attenuation rate distributions of each sound source grid point to a microphone, which are required for beamforming, using the reduced-order model with sparse sampling for the acceleration of the computation. Furthermore, it should be noted that the data-driven sparse sampling has recently been applied to the online temperature fields reconstruction of steam turbine casing \citep{jiang2022online}, to the fast computation of inverse transient analysis for pipeline condition assessment \citep{wang2022fast}, and to the improvement of signal-to-noise ratio in noisy measurement \citep{inoue2021data}.

For the optimization of the sparse sensor positions, pressure-sensitive paint (PSP) is useful for the measurement of the surface pressure distributions, which are the training data. PSP is a type of fluorescent paint that produces luminescence according to the partial pressure of oxygen when irradiated with excitation light. Utilizing this feature, the pressure distribution can be obtained by exciting the PSP applied to the model and measuring it with a charge-coupled device (CCD) camera \citep{liu2021pressure}. This technique is actively used in the aeronautical field \citep[e.g.][]{merienne2013transonic,watkins2016measuring,sugioka2018experimental,masini2020analysis,uchida2021analysis}. The low-pressure conditions \citep{anyoji2015effects,nagata2020optimum,nagata2020experimental,kasai2021frequency} and rarefied and/or microflows \citep{niimi2005application,huang2015applications,huang2020application} are also scopes of application. Since the PSP is an absolute pressure sensor, it is generally difficult to accurately measure the pressure by PSP at low- speed conditions such as automotive aerodynamics in which pressure changes are small. However, low-speed PSP measurement has been becoming possible through various efforts such as temperature correction of PSP using thermocouples \citep{yamashita2007pressure}, optimization of images acquisition timing \citep{brown2000low,bell2004applications,mitsuo2005temperature} and number \citep{yorita2012development}, and use of new paint with less temperature dependence \citep{gouterman2004dual}. Therefore, research applying PSP measurement to automobile models is also being conducted \citep[e.g.][]{yamashita2007pressure,aider2001pressure,gouterman2004dual,engler2002application}. 
Furthermore, the recent development of the PSP and data science techniques, such as fast Fourier transform (FFT), proper orthogonal decomposition (POD) \citep{berkooz1993proper, taira2017modal}, and sparse sampling, have made low-speed unsteady PSP measurements possible. For example, \cite{egami2020ruthenium} developed a novel ruthenium complex-based fast-PSP and succeeded in obtaining time-series data of pressure fluctuation with high accuracy even at a mean velocity of 20~m/s. \cite{peng2016fast} applied POD analysis to low-speed unsteady PSP data and confirmed  POD analysis could effectively remove noise. \cite{wen2018data} developed a data mining approach based on compressed data fusion and succeeded in recovering the unsteady pressure field induced by a cylinder wake flow at low speed from the highly noisy PSP data. In addition to these, there are many other studies that have taken on the challenge of low-speed unsteady PSP measurements \citep[e.g.][]{inoue2021data,sugioka2019unsteady,noda2018detection,ozawa2019cross}.

In this study, the Ahmed model \citep{ahmed1984some}, also known as the simple automobile model, is used to validate the proposed framework. The flow around an Ahmed model is representative of the flow around the actual vehicle and is a good reference for the investigation of the aerodynamic characteristics of the actual vehicle \citep{yu2021recent}. There are a number of studies on the elucidation of the flow structure around an automobile using the Ahmed model, both experimental \cite[e.g.][]{ahmed1984some,thacker2012effects,tunay2016computational} and numerical \cite[e.g.][]{krajnovic2005influence,guilmineau2008computational,corallo2015effect,keogh2016aerodynamic}. The effects of vehicle spacing \citep{watkins2008effect} and overtaking \citep{uystepruyst2013numerical} on aerodynamics have also been investigated with the Ahmed model. In this study, time-averaged pressure distribution data on an Ahmed model at various yaw angles are obtained by the PSP measurement as training data, and sparse sensor position optimization is performed for the yaw angle estimation and the pressure distribution reconstruction. Three different sensor position optimization algorithms based on the greedy algorithm are applied, and the sensor positions selected by each algorithm and the relationship between the number of sensors and the estimation error of the pressure distribution and the yaw angle are discussed.

\section{Computational Algorithm}
\begin{figure}[htbp]
    \includegraphics[width=1.0\textwidth]{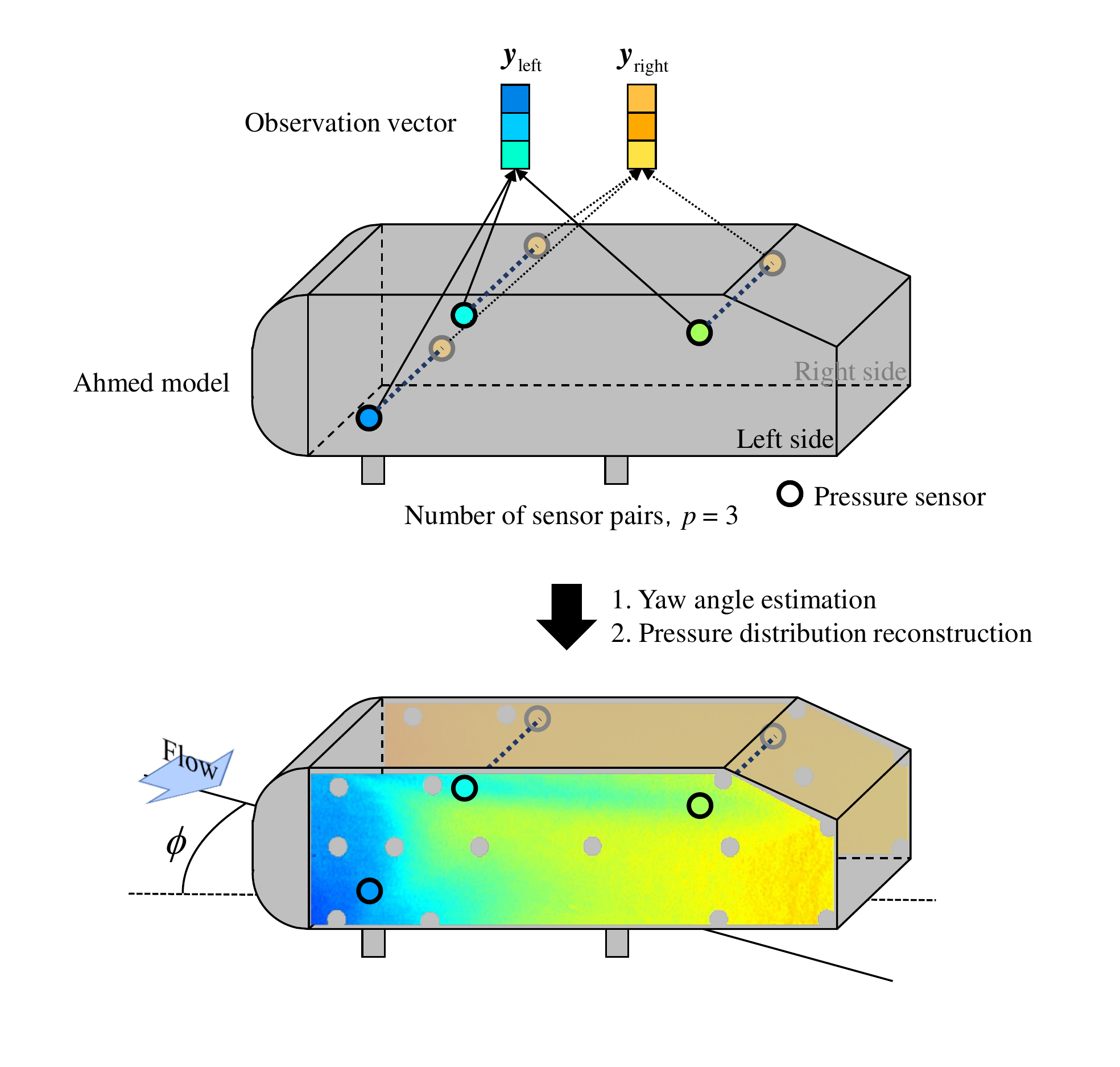}
    \caption{Schematic of the yaw angle estimation and the pressure
distribution reconstruction}
    \label{fig:overview}
\end{figure}

This study uses the time-averaged pressure coefficient $C_P$ distribution data on the left and right sides of the Ahmed model at various yaw angles. Here, a data vector of pixel units, converted from the luminescence intensity captured by PSP images, is represented as the $C_P$ distribution. Figure \ref{fig:overview} shows the schematic of the yaw angle estimation and pressure distribution reconstruction. The $p$-sparse pressure sensor pairs are installed at the same position on the left and right sides of the model. Here, the symmetric sensors in the left and right sides of the model were assumed, and therefore, the term ``sensor pair'' represents the sensors symmetrically placed on the left and right sides of the body in this study. The yaw angle is estimated and the pressure distribution is reconstructed from the observed values $\bm{y}$ from the sensors. The optimization of the placement of the sparse pressure sensors corresponds to the problem of choosing the pixel to be observed. Here, the number of sensor pairs is reduced by using a common sensor for the yaw angle estimation and the pressure distribution reconstruction.

The yaw angle is positive counterclockwise. Therefore, when the yaw angle is positive, the left side is leeward.
When PSP measurements are conducted on the left side of the model for each yaw angle from $-\varphi_l$ to $\varphi_l$, the vector that stores the yaw angle $\bm{\phi}\in\mathbb{R}^{m}$ is defined as follows:
\begin{align}
\begin{split}
    \bm{\phi}
    &=\begin{bmatrix}
    \phi_1 & \phi_2 & \cdots & \phi_{m}
    \end{bmatrix}
    ^{\mathsf T}\\
    &=\begin{bmatrix}
    -\varphi_l & \cdots & -\varphi_1  & 0 & \varphi_1 & \cdots & \varphi_l
    \end{bmatrix}
    ^{\mathsf T},
    \label{eq:phi}
\end{split}
\end{align}
where $l=\left(m-1\right)/2$ and $\phi_{l+1}=\varphi_0=0$. Here, the superscript $\mathsf{T}$ represents the transpose of the matrix.
The pressure data matrix on the left side $\bm{X}_{\rm left}\in\mathbb{R}^{n\times m}$ is defined as follows:
\begin{align}
    \bm{X}_{\rm left}=
    \begin{bmatrix}
    \bm{x}_1 & \bm{x}_2 & \cdots & \bm{x}_m
    \end{bmatrix},
    \label{eq:xleft}
\end{align}
where $\bm{x}_j\in\mathbb{R}^{n}$ is a column vector of the time-averaged pressure distribution on the left side at an arbitrary yaw angle. The subscript number $j$ of $\phi_j$ corresponds to that of $\bm{x}_j$.

Due to the assumption of the symmetry of the left and right sides, the pressure data matrix of the right side $\bm{X}_{\rm right}\in\mathbb{R}^{n\times m}$ can be written as:
\begin{align}
    \bm{X}_{\rm right}=
    \begin{bmatrix}
    \bm{x}_m & \bm{x}_{m-1} & \cdots & \bm{x}_1    
    \end{bmatrix}.
    \label{eq:xright}
\end{align}
Hysteresis is not considered in this method.

The sensor position is determined by the sensor position optimization algorithm using $\bm{\phi}, \bm{X}_{\rm left}, \bm{X}_{\rm right}$ as the training data. The $k$-th sensor pair defines the sensor position vector $\bm{h}_{k}\in\mathbb{R}^{n}$ using the sensor index $i_k$ as follows:
\begin{align}
    h_k\left(i\right)= \left\{
    \begin{array}{ll}
         &  1 \;\left(i=i_k\right)\\
         &  0 \;\left(i \neq i_k\right)
    \end{array}
    \right.
    \label{eq:hk}
\end{align}
Here, $\bm{h}_{\rm k}$ is a sparse column vector where the element corresponding to the $k$-th sensor position is 1 and the other elements are 0. In this case, the sensor position matrix $\bm{H}\in\mathbb{R}^{p\times n}$ and the sensor-pair position matrix extended to symmetric layout $\bm{\Pi}\in\mathbb{R}^{2p\times 2n}$ can be written as follows:
\begin{align}
    \bm{H}&=
    \begin{bmatrix}
    \bm{h}_1 & \bm{h}_2 & \cdots & \bm{h}_p
    \end{bmatrix}
    ^{\textsf T},
    \label{eq:H}\\
    \bm{\Pi}&=
    \begin{bmatrix}
    \bm{h}_1 & \bm{0} & \bm{h}_2 & \bm{0} & \cdots & \bm{h}_p & \bm{0} \\
    \bm{0} & \bm{h}_1 & \bm{0} & \bm{h}_2 & \cdots  & \bm{0} & \bm{h}_p
    \end{bmatrix}
    ^{\textsf T}.
    \label{eq:Pi}
\end{align}
Sensor position optimization is performed using the vectors and the matrices defined above, and thus, the yaw angle estimation and the pressure distribution reconstruction are conducted.

\subsection{Estimation method}
In this study, the yaw angle estimation and the pressure distribution reconstruction are based on the linear regression and the least-squares estimation of the POD coefficients, respectively. This study applied the POD analysis to the pressure data matrix $\bm{X}$ and derived a low-dimensional model of the change in pressure distribution with respect to the change in yaw angle. Then, the POD coefficients were estimated from the pressure data obtained by pressure sensors for the pressure distribution reconstruction.
Since the time-averaged PSP measurement noise can be assumed to be sufficiently small, least-squares regression is available for estimation. It should be noted that the estimation based on least-squares regression may not be appropriate when the measurement includes outliers and/or is strongly contaminated. In such cases, it is necessary to apply a method such as the $\ell_1$ regression or the compressed sensing \citep{wen2018data}.

\subsubsection{Yaw angle estimation}
The yaw angle estimation is performed using the linear regression with the differential pressures of the left and right sensor pairs. 
\begin{align}
    \widehat{\phi}=\bm{y}_{\rm dif}^{\mathsf T}\bm{\beta},
    \label{eq:phiest}
\end{align}
where $\widehat{\phi}\in\mathbb{R}^{1}$, $\bm{y}_{\mathrm{dif}}\in\mathbb{R}^{p}$, and $\bm{\beta}\in\mathbb{R}^{p}$ are the estimated yaw angle, the left and right differential pressure observation vector, and the coefficient vector, respectively. At an arbitrary yaw angle, the $\bm{y}_{\mathrm{dif},j}$ can be written as:
\begin{align}
    \begin{split}
        \bm{y}_{\mathrm{dif},j}
        &=\bm{H}\left(
        \bm{x}_{\mathrm{left},j}-
        \bm{x}_{\mathrm{right},j}
        \right)\\
        &=\bm{H}\left(
        \bm{x}_{j}-
        \bm{x}_{m-j+1}
        \right).
    \end{split}
    \label{eq:ydif}
\end{align}

The coefficient vector $ \bm{\beta} $ can be obtained by the least-squares estimation using the training data, as shown in Eq.~\eqref{eq:beta}. At $ \phi=0 $, the left and right differential pressures are zero in this method, so the regression line is a straight line passing through the origin.
\begin{align}
    \bm{X}_{\mathrm{dif}}&=
    \bm{X}_{\mathrm{left}}-
    \bm{X}_{\mathrm{right}},
    \label{eq:Xdif}\\
    \bm{Y}_{\mathrm{dif}}&=
    \bm{H}\bm{X}_{\mathrm{dif}},
    \label{eq:Ydif}\\
    \begin{split}
    \bm{\beta}
    &=\left(\bm{Y}_{\mathrm{dif}}^{\mathsf{T}}\right)^+\bm{\phi}\\
    &=\left\{
    \begin{array}{ll}
         &  \bm{Y}_{\mathrm{dif}}
         \left(\bm{Y}_{\mathrm{dif}}^{\mathsf{T}}\bm{Y}_{\mathrm{dif}}\right)^{-1}
         \bm{\phi} \;\left(p\leq m\right)\\
         &  \left(\bm{Y}_{\mathrm{dif}}\bm{Y}_{\mathrm{dif}}^{\mathsf{T}}\right)^{-1}
         \bm{Y}_{\mathrm{dif}}\bm{\phi} \;\left(p>m\right)
    \end{array}
    \right.    
    \end{split}
    \label{eq:beta}
\end{align} 
where the superscript $+$ represents the Moore--Penrose inverse of the matrix.

\subsubsection{Pressure distribution reconstruction}
The reconstruction method based on a tailored basis proposed by \cite{manohar2018data} is employed for the reconstruction of the pressure distribution. The tailored basis is generated by data-driven techniques. The POD analysis was adopted to obtain the tailored basis of the pressure field, the same as the original work, and the obtained basis was used for pressure field reconstruction.
Here, POD is one of the effective methods for extracting significant modes from high-dimensional data. If the data can be effectively expressed by a limited number of POD modes, limited sensors placed at appropriate positions will give the approximated full state information by estimating the POD mode coefficients. This idea has been extended to a further generalized form \citep{clark2018greedy,clark2020multi,clark2020sensor,manohar2021optimal,saito2021determinant,yamada2021fast,yamada2021greedy}.

The data matrix $\bm{X}_{\mathrm{both}}\in\mathbb{R}^{2n\times m}$, which stores the left and right side pressure distribution data, is created by vertically combining $\bm{X}_{\mathrm{left}}$ and $\bm{X}_{\mathrm{right}}$ as follows:
\begin{equation}
    \bm{X}_{\mathrm{both}}=
    \begin{bmatrix}
    \bm{X}_{\mathrm{left}} \\
    \bm{X}_{\mathrm{right}}
    \end{bmatrix}.
    \label{eq:Xboth}    
\end{equation}

The data matrix $\widetilde{\bm{X}}_\mathrm{both}$ is created by subtracting the averaged pressure distribution data at different yaw angles $\overline{\bm{x}}_\mathrm{both}$ from each column of the matrix $\bm{X}_{\mathrm{both}}$. Then, POD modes are calculated by applying singular value decomposition to $\widetilde{\bm{X}}_\mathrm{both}$.
\begin{align}
    \overline{\bm{x}}_\mathrm{both}&=
    \cfrac{1}{m} \sum_{j=1}^m \bm{x}_{\mathrm{both},j}\\
    \begin{split}
        \widetilde{\bm{X}}_\mathrm{both}
        &=\bm{X}_{\mathrm{both}}-
        \begin{bmatrix}
            1 & \cdots & 1
        \end{bmatrix}
        \overline{\bm{x}}_{\mathrm{both}}\\
        &=\begin{bmatrix}
        \bm{U}_{\mathrm{left}} \\
        \bm{U}_{\mathrm{right}}
        \end{bmatrix}\bm{\Sigma V}^{\mathsf{T}}\\
        &=\bm{U Z}\\
        &\approx\bm{U}_{1:r}\bm{Z}_{1:r}
    \end{split}
    \label{eq:Xboth2}
\end{align}
Here, $\bm{U}, \bm{\Sigma}, \bm{V}$, and $\bm{Z}$ are the spatial POD modes, the singular values, the yaw angle POD modes, and the POD coefficients, respectively. The leading-$r$ modes are selected in descending order of singular value and used for the construction of the low-dimensional model. The pressure distribution is reconstructed by the least-squares estimation of the POD coefficients from the observed sensor values, as shown below:
\begin{align}
    \bm{D}&=
    \bm{\Pi U}_{1:r},
    \label{eq:D}\\
    \begin{split}
        \widehat{\bm{z}}&=
        \bm{D}^+\bm{y}_{\mathrm{both}}\\
        &=\left\{
        \begin{array}{ll}
             &  \bm{D}^{\mathsf{T}}
             \left(\bm{D}\bm{D}^{\mathsf{T}}\right)^{-1}
             \bm{y}_{\mathrm{both}}
             \;\left(2k\leq r\right)\\
             &  \left(\bm{D}^{\mathsf{T}}\bm{D}\right)^{-1}
             \bm{D}^{\mathsf{T}}
             \bm{y}_{\mathrm{both}}
             \;\left(2k>r\right),
        \end{array}
        \right.
    \end{split}
    \label{eq:zest}\\
    \widehat{\bm{x}}_\mathrm{both}&=
    \bm{U}_{1:r} \widehat{\bm{z}} + \overline{\bm{x}}_{\mathrm{both}},
    \label{eq:xbothest}
\end{align}
where $ \bm{y}_\mathrm{both}\in\mathbb{R}^{2p} $ is the observation vector. At an arbitrary yaw angle, the $ \bm{y}_{\mathrm{both},j} $ can be written as:
\begin{align}
    \bm{y}_{\mathrm{both},j}=
    \bm{\Pi}
    \left(
    \bm{x}_{\mathrm{both},j}-\overline{\bm{x}}_{\mathrm{both}}
    \right).
    \label{eq:yboth}
\end{align}

\subsection{Sensor position optimization algorithm}
In this study, three types of algorithms based on the greedy algorithm are used for sensor optimization. 
There are several other optimization methods \citep{joshi2009sensor,nonomura2021randomized,nagata2021data} rather than greedy algorithms, but the greedy algorithm is adopted in this study for computational efficiency. 
The greedy algorithm divides the problem into local subproblems and sequentially incorporates the solution that maximizes or minimizes the objective function in each of the subproblems. Although it does not necessarily give the global optimal solution, it can significantly reduce the computational cost compared to the brute-force search that verifies all combinations. Each algorithm is explained in the following sections.

\subsubsection{Orthogonal matching pursuit}
The orthogonal matching pursuit (OMP) algorithm is an algorithm that iterates a minimization problem with 1-sparse vectors as a subproblem for solving $\ell_0$ norm optimization problems. By using the idea of orthogonality, once an index is selected, it is not selected again \citep{pati1993orthogonal, davis1994adaptive}. In this paper, by the application of the OMP algorithm, the suboptimum sensor position for the yaw angle estimation was searched for by using the $\ell_2$ norm minimization of the estimated yaw angle residual as the objective function.
\begin{align}
    \mathrm{minimize}
    \left\| 
    \left(\bm{HX}_{\mathrm{dif}}\right)^{\mathsf{T}}\bm{\beta} - \bm{\phi} 
    \right\|_2^2 
\end{align}

The OMP algorithm is summarized in Alg.~\ref{alg:OMP}. In the OMP algorithm, one sensor pair is selected for each iteration. It should be again noted that the symmetric sensor positions in the left and right sides of an automobile are assumed in this study, and therefore, the choice in one index in the vector corresponds to the choice in the sensor positions of one sensor pair. Here, $x_{\mathrm{dif},i,j}$ and $\bm{r}_k$ are the $(i,j)$ element of $\bm{X}_{\mathrm{dif}}$ and the $k$-th residual, respectively.

\begin{algorithm}[H]
    \caption{Orthogonal Matching Pursuit Algorithm}
    \label{alg:OMP}
    \begin{algorithmic}[1]
    \STATE Set $\bm{r}_0 = \bm{\phi}$, $\bm{H}_0 = $ {\O}
    \FOR{$k=1$ to $p$}
    \STATE $\bm{x}_{\mathrm{dif,} i} = \begin{bmatrix}
            x_{\mathrm{dif,} i,1} & x_{\mathrm{dif,} i,2} & \cdots & x_{\mathrm{dif,} i,m}
            \end{bmatrix}$
    \STATE $i_k \leftarrow 
            \argmax_{i} \cfrac
            {\left\langle \bm{x}_{\mathrm{dif,} i}^\mathsf{T} , \,\bm{r}_{k-1} \right\rangle^2}
            {\left\| \bm{x}_{\mathrm{dif,} i}^\mathsf{T} \right\|_2^2}$ 
    \STATE $\bm{h}_k\left(i_k\right) = 1$
    \STATE $\bm{H}_k = 
            \left[\bm{H}_{k-1}^\mathsf{T}, \bm{h}_k\right]^\mathsf{T}$
    \STATE $\bm{\beta}_k = \left(\left(\bm{H}_k \bm{X}_{\mathrm{dif}}\right)
            ^{\mathsf{T}}\right)^+ \bm{r}_{k-1}$
    \STATE $\bm{r}_k = \bm{\phi} - 
            \left(\bm{H}_k \bm{X}_{\mathrm{dif}}\right)^{\mathsf{T}} \bm{\beta}_k$
    \ENDFOR
    \end{algorithmic}
\end{algorithm}

\subsubsection{D-optimality-based greedy algorithm for vector-measurement sensors}
The D-optimality-based greedy algorithm for vector-measurement sensors (DG-vector) \citep{saito2020data,saito2021data} is an algorithm based on the D-optimal design of experiments that maximize the determinant of the Fisher information matrix ($\bm{DD}^{\mathsf{T}}$ or $\bm{D}^{\mathsf{T}}\bm{D}$ ). Here, the sensor placement technique \citep{saito2020data,saito2021data} for the vector-component sensor is utilized in the present study because the choice of one index of $\bm{h_k}$ corresponds to the choice of two sensors of a sensor pair owing to the symmetric sensor location in the left and right sides of the body. 
\begin{align}
    \mathrm{maximize}
    \left\{
    \begin{array}{ll}
         &  \det\left(\bm{DD}^{\mathsf{T}}\right)
         \;\left(2k\leq r\right)\\
         &  \det\left(\bm{D}^{\mathsf{T}}\bm{D}\right)
         \;\left(2k>r\right)
    \end{array}
    \right.
    \label{eq:DG}
\end{align}
By maximizing $\bm{DD}^{\mathsf{T}}$ or $\bm{D}^{\mathsf{T}}\bm{D}$, the pseudo-inverse of $\bm{D}$ can be obtained stably, and the error in pressure distribution reconstruction can be minimized.

The DG-vector algorithm is summarized in Alg.~\ref{alg:DG}. In the DG-vector algorithm, sensor pairs are selected one by one in each step. Here, $u_{i,j}$ is the $(i,j)$ element of $\bm{U}$, and $\epsilon$ is a sufficiently small value \citep{saito2021determinant}, which was empirically set to $10^{-10}$ in this paper. 

\begin{algorithm}[H]
    \caption{D-optimality-based Greedy Algorithm for Vector-Measurement Sensors}
    \label{alg:DG}
    \begin{algorithmic}[1]
    \STATE Set $\bm{D}_0^{\mathsf{T}}\bm{D}_0 = \epsilon\bm{I}$
    \FOR{$k=1$ to $p$}
    \STATE $\bm{W}_i = \begin{bmatrix}
                {u}_{i,1} & {u}_{i,2} & \cdots & {u}_{i,r}\\
                {u}_{n+i,1} & {u}_{n+i,2} & \cdots & {u}_{n+i,r}
            \end{bmatrix}$
    \STATE $i_k \leftarrow 
            \argmax_{i} \det
            \left(\bm{I} + \bm{W}_i 
                \left(\bm{D}_{k-1}^{\mathsf{T}}\bm{D}_{k-1}\right)^{-1}
                \bm{W}_i^{\mathsf{T}}
            \right)$
    \STATE $\left(\bm{D}_{k}^{\mathsf{T}}\bm{D}_{k}\right)^{-1}$\\
    \STATE $\quad \leftarrow
                \left(\bm{D}_{k-1}^{\mathsf{T}}\bm{D}_{k-1}\right)^{-1}$\\
    \STATE $\qquad \left(\bm{I} - \bm{W}_{i_k}^{\mathsf{T}}
                \left(\bm{I} + \bm{W}_{i_k}
                    \left(\bm{D}_{k-1}^{\mathsf{T}}\bm{D}_{k-1}\right)^{-1}
                    \bm{W}_{i_k}^{\mathsf{T}}
                \right)^{-1} \bm{W}_{i_k}
                \left(\bm{D}_{k-1}^{\mathsf{T}}\bm{D}_{k-1}\right)^{-1}
            \right)$
    \STATE $\bm{h}_k\left(i_k\right) = 1$
    \STATE $\begin{bmatrix}
                {u}_{i_k,1} & {u}_{i_k,2} & \cdots & {u}_{i_k,r} 
            \end{bmatrix} \leftarrow
            \begin{bmatrix}
                0 & 0 & \cdots & 0
            \end{bmatrix}$
    \STATE $\begin{bmatrix}
                {u}_{n+i_k,1} & {u}_{n+i_k,2} & \cdots & {u}_{n+i_k,r} 
            \end{bmatrix} \leftarrow
            \begin{bmatrix}
                0 & 0 & \cdots & 0
            \end{bmatrix}$
    \ENDFOR
    \end{algorithmic}
\end{algorithm}

\subsubsection{Hybrid algorithm}
In principle, the OMP and DG-vector algorithms are sensor position optimization algorithms suitable for the yaw angle estimation and pressure distribution reconstruction, respectively. Therefore, the sensor pairs selected by these algorithms may not be suitable for both yaw angle estimation and pressure distribution reconstruction. In addition, the OMP algorithm selects the sensor pairs that minimize the residual of the estimated yaw angle at each step and are prone to overfitting when the number of sensor pairs is large. Therefore, switching the algorithm to the DG-vector algorithm when the OMP algorithm overfits can be expected to further improve performance. Thus, we tried the OMP-DG hybrid algorithm (hereafter, the hybrid algorithm), which uses the OMP algorithm to select from first to $q$-th sensor pairs and the DG-vector algorithm to select from ($q+1$)-th to $p$-th sensor pairs in this study. Here, $q$ represents the number of sensor pairs to which the OMP algorithm overfits, and is determined based on the selected sensor pair positions and/or the relationship between the number of sensor pairs and the estimation error. 

The hybrid algorithm is summarized in Alg.~\ref{alg:Hybrid}. The hybrid algorithm uses the index of $q$ sensor pairs selected by the OMP algorithm to determine the next sensors by the DG-vector algorithm. Therefore, the position of the first sensor pair selected by the DG-vector algorithm and the position of the $(q+1)$-th sensor pair selected by the hybrid algorithm are not the same. 

In this way, a combination of the greedy algorithms such as the OMP and DG-vector algorithms, which determine one sensor-pair position at each step, is straightforward. The performance improvements can be expected by considering the nature of each algorithm and combining them appropriately.

\begin{algorithm}[H]
    \caption{OMP-DG Hybrid Algorithm}
    \label{alg:Hybrid}
    \begin{algorithmic}[1]
    \STATE Obtain $\bm{H}_q$ and $\bm{\Pi}_q$ based on Alg.~\ref{alg:OMP}
    \STATE $\bm{D}_q^{\mathsf{T}}\bm{D}_q 
            = \left(\bm{\Pi}_q\bm{U}_r\right)^{\mathsf{T}}\bm{\Pi}_q\bm{U}_r$
    \STATE Obtain $\bm{H}_p$ and $\bm{\Pi}_p$ based on Alg.~\ref{alg:DG} from $k=q+1$
    \end{algorithmic}
\end{algorithm}

\section{Pressure Data Acquisition Experiment and Data Processing}
In this study, the pressure distribution data on the left side of the Ahmed model were acquired using PSP. The intensity-based method was used to measure pressure distributions by utilizing the PSP character that the change in luminescence intensity depends on the pressure, as described by the Stern-Volmer equation:
\begin{align}
    \cfrac{I_{\mathrm{ref}}}{I}=
    A+B
    \cfrac{P}{P_{\mathrm{ref}}},
    \label{eq:SternVolmer}
\end{align}
where $I$ and $I_\mathrm{ref}$ are the luminescence intensity of a run image at the wind-on condition and a reference image at the wind-off condition, and $P$ and $P_\mathrm{ref}$ are the pressure under the run and reference conditions, respectively. The constants $A$ and $B$ are the Stern--Volmer coefficients.

\subsection{Experimental method}
\begin{figure}[!htbp]
    \centering
    \includegraphics[width=0.8\textwidth]{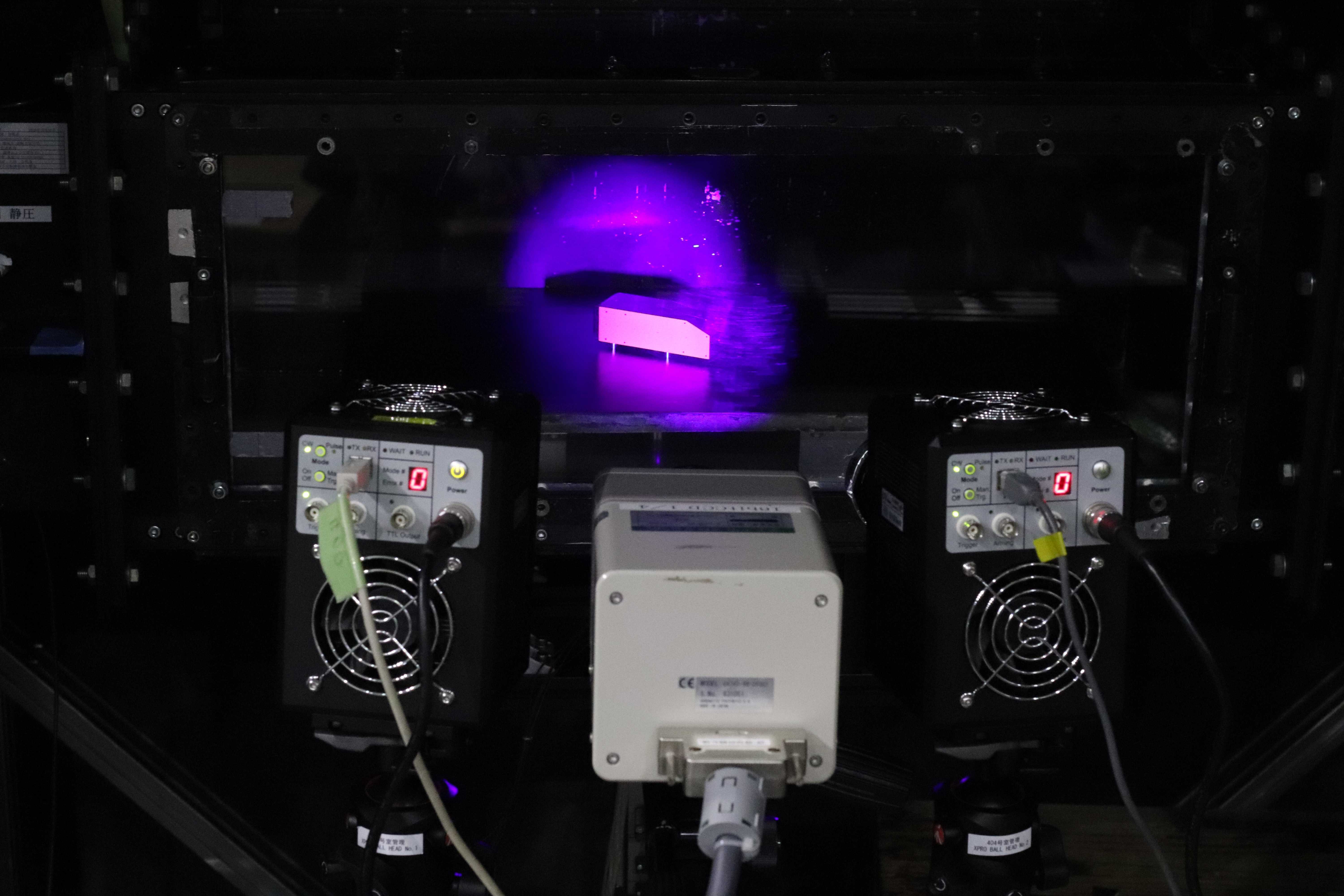}\\
    \includegraphics[width=1.0\textwidth]{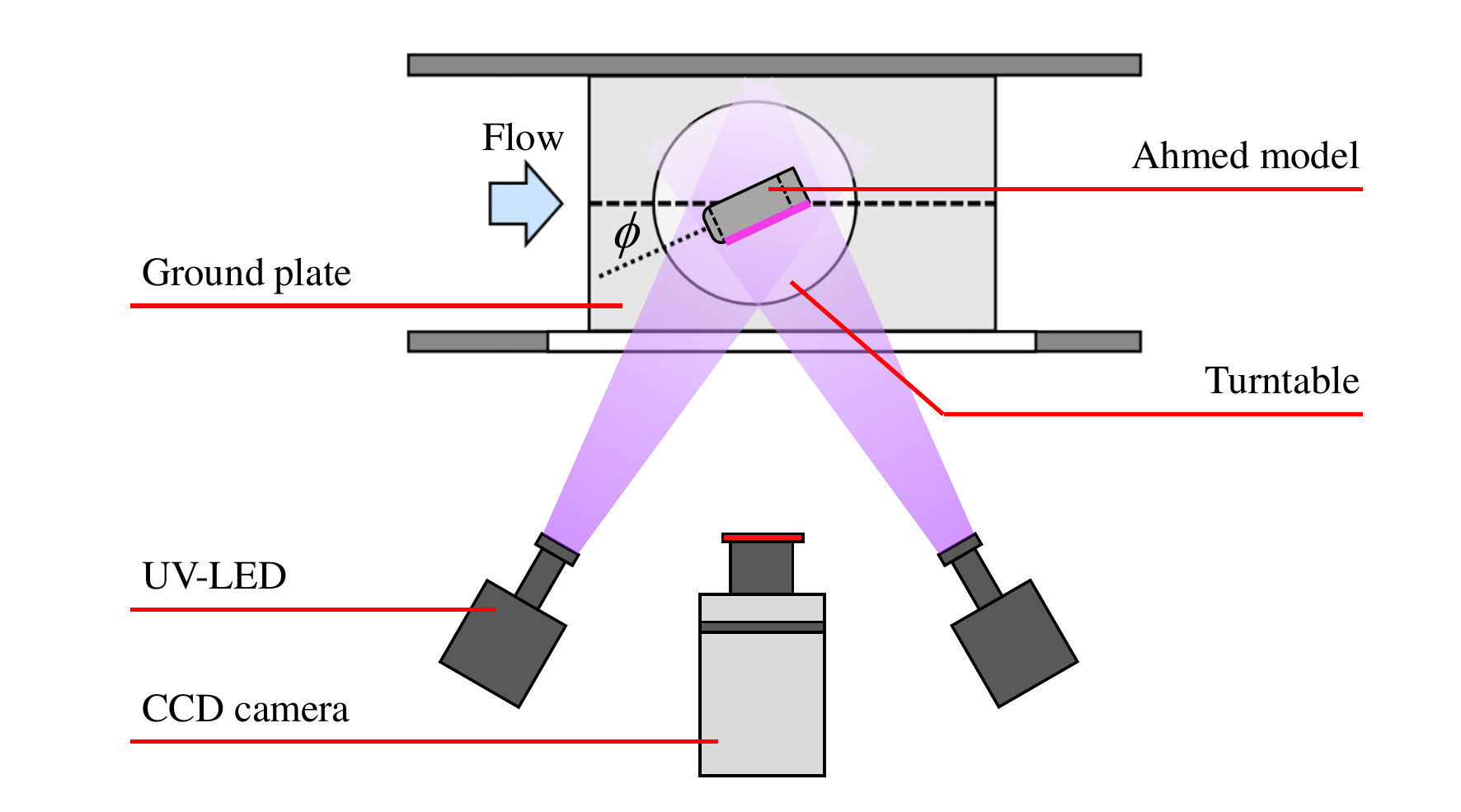}
    \caption{Picture and schematic of the PSP measurement system}
    \label{fig:experiment}
\end{figure}

\begin{figure}[!htbp]
    \centering
    \includegraphics[width=1.0\textwidth]{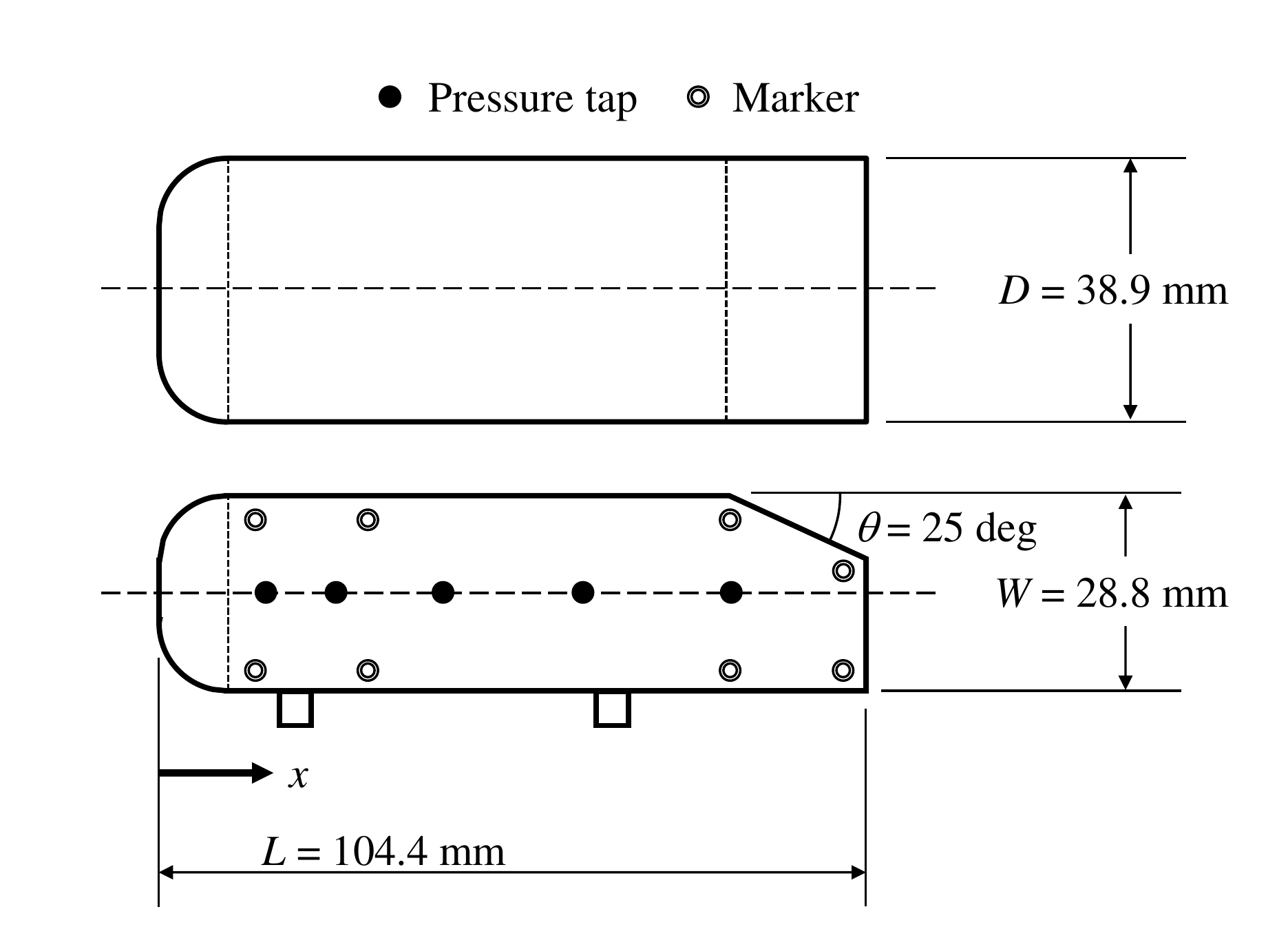}
    \caption{Schematic of the 1/10 scaled Ahmed model \citep{ahmed1984some}}
    \label{fig:AhmedModel}
\end{figure}

The present experimental setup was similar to that of the previous study by \cite{yamashita2007pressure}. A picture and schematic of the PSP measurement system are shown in Fig.~\ref{fig:experiment}. This experiment was conducted in Tohoku-University Basic Aerodynamic Research Wind Tunnel (T-BART). The model used in this study was a 1/10 scaled Ahmed model \citep{ahmed1984some} and the schematic of the model is shown in Fig.~\ref{fig:AhmedModel}. The length $L$, the width $W$, and the height $H$ of the model were 104.4~mm, 38.9~mm, and 28.8~mm, respectively, and the rear slant angle $\theta$ is 25~deg. On the left side of the model, there were five pressure taps  ($x/L = $0.15, 0.25, 0.4 0.6, and 0.81), which were connected to the pressure transducers (30~INCH-D1-4V-MINI,  Amphenol All Sensors) through tubes, to compare with PSP results. UniFIB and FIB Basecoat (Innovative Scientific Solutions Incorporated, ISSI) were painted on the left side of the model. UniFIB is a commercially available PSP, and FIB Basecoat is an undercoat for UniFIB. In addition, eight black dots (markers) were put on the surface of the model for image registration between the reference and run images. The model was installed on a ground plate (length: 600~mm, width: 296~mm, thickness: 15~mm) representing the ground. The leading edge of the ground plate was an elliptical shape to prevent the leading edge separation. The Ahmed model was mounted on a rotary table located at the center of the ground plate, and the model yaw angle was changed by the rotary table. Two ultra-violet light-emitting diodes (LED) devices (IL-106, HARDsoft) and a 16-bit CCD camera (C4742-98, Hamamatsu Photonics K.K.) were used as excitation light sources, and an emission detector, respectively. A camera lens (Nikkor 50~mm f/1.4, Nikon) was attached to the camera. Additionally, a $650\pm20$~nm band-pass filter (PB0124, Asahi Spectra Co., Ltd) was attached to the camera lens to cut off undesirable light, such as the excitation lights from LEDs.

The freestream velocity was set to 50~m/s, and the Reynolds number $Re_L$ based on the total length of the model was $3.4\times10^5$. The wind tunnel was stopped immediately after the run images were acquired, and the reference images and dark images were acquired. The number of acquired images for the run, reference, and dark images was 16. The yaw angle of the model was changed by 1.0~deg in the range from $-25$~deg to 25~deg, and the PSP measurements were taken for 51 cases ($m=51$).

\subsection{Analysis method}
First, the run, reference, and dark images were ensemble-averaged, respectively, and then the dark images were subtracted from the run and reference images. After the image registration of the run and reference images, the luminescence intensity ratios were calculated. Then, the $C_P$ distributions were calculated using Eq.~\eqref{eq:SternVolmer}. Here, the Stern-Volmer coefficients $A$ and $B$ were calculated using the in-situ calibration method based on the luminescence intensity ratio around the pressure taps and the pressure value obtained at the pressure taps.
The Wiener filter with a window range of $3 \times 3$ was applied to $C_P$ distributions to remove the noise.
The matrix $\bm{X}_{\mathrm{left}}$ was created from the $C_P$ data on the left side of the model acquired by the PSP measurements. The robust principal component analysis (robust PCA) \citep{candes2011robust} was applied to $\bm{X}_{\mathrm{left}}$ to remove the noise. The matrix $\bm{X}_{\mathrm{right}}$ was also created by flipping $\bm{X}_{\mathrm{left}}$ with assuming symmetricity. The number of pixels $n$ in the pressure distribution was 79,931. 

\subsection{Verification Method}
In this study, two methods for verification were adopted. In the first validation (the validation I), the performance of the algorithms was evaluated by using training data as test data. In the second validation (the validation II), the generalization performance of the algorithms and low-rank model was evaluated by separating the training data and test data using leave-one-out cross-validation \citep{brunton2019data}. The validation II is summarized in Alg.~\ref{alg:ValidationII}. Since $\bm{\phi}$, $\bm{X}_{\mathrm{dif}}$, and $\bm{X}_{\mathrm{both}}$ have symmetry between the negative and positive yaw angles, these validations were conducted only in the range from 0~deg to $\varphi_l~(=25)$~deg.
\begin{algorithm}[H]
    \caption{Validation II (leave-one-out cross-validation)}
    \label{alg:ValidationII}
    \begin{algorithmic}[1]
    \FOR{$j=(m+1)/2$ to $m$}
    \STATE Create the data matrices for cross-validation $\bm{\phi}^{cv}$, $\bm{X}_{\mathrm{dif}}^{cv}$, and $\bm{X}_{\mathrm{both}}^{cv}$ by removing the $j$-th and $(m-j+1)$-th elements from $\bm{\phi}$, $\bm{X}_{\mathrm{dif}}$, and $\bm{X}_{\mathrm{both}}$, respectively.
    \STATE Using $\bm{\phi}^{cv}$, $\bm{X}_{\mathrm{dif}}^{cv}$, and $\bm{X}_{\mathrm{both}}^{cv}$, construct the estimation model and optimize the sensor placement.
    \STATE Using $\phi_j$, $\bm{x}_{\mathrm{dif},j}$, and $\bm{x}_{\mathrm{both},j}$, test the constructed estimation model and selected sensor placement.
    \ENDFOR
    \end{algorithmic}
\end{algorithm}

The errors of the yaw angle estimation and the pressure distribution reconstruction were calculated using the following equations. Equations~\eqref{eq:error1} and \eqref{eq:error3} are used for the validation I, and Eqs.~\eqref{eq:error2} and \eqref{eq:error4} are used for the validation II, respectively.

\begin{align}
    e_{\phi}&=
    \cfrac
    {\left\|\widehat{\bm{\phi}} - \bm{\phi} \right\|_2}
    %{\sqrt{l+1}}
    {\sqrt{\left(m+1\right)/2}}
    \label{eq:error1}\\
    e_{\phi,j}&=
    \left|\widehat{\phi}_j - \phi_j \right|
    \label{eq:error2}\\
    e_{P}&=
    \cfrac
    {\left\|\widehat{\bm{X}}_\mathrm{both} - \bm{X}_\mathrm{both} \right\|_\mathrm{F}}
    %{\sqrt{2n\left(l+1\right)}}
    {\sqrt{2n\cdot\left(m+1\right)/2}}
    \label{eq:error3}\\
    e_{P,j}&=
    \cfrac
    {\left\|\widehat{\bm{x}}_{\mathrm{both},j} - \bm{x}_{\mathrm{both},j} \right\|_2}
    {\sqrt{2n}}
    \label{eq:error4}
\end{align}

\section{Results and Discussion}
\subsection{Basic characteristics of pressure distribution}
\begin{figure}[!htbp]
    \centering
    \subfigure[-25 deg]{
    \includegraphics[width=0.32\textwidth]{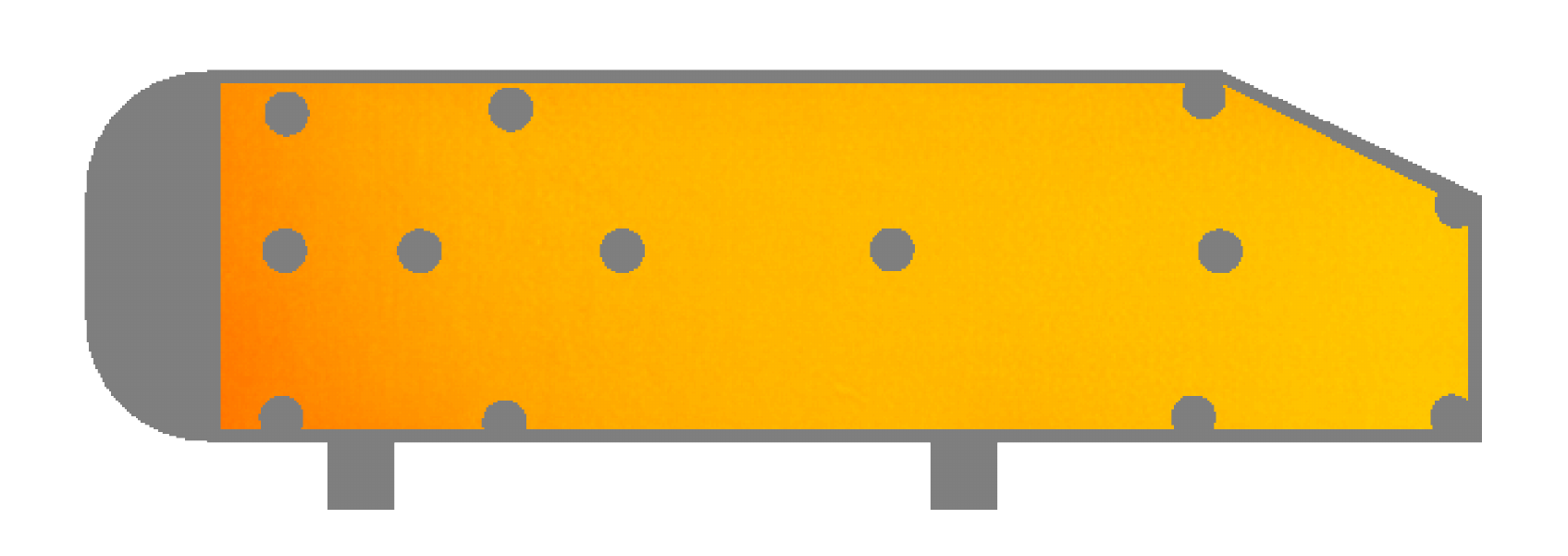}}
    \subfigure[-20 deg]{
    \includegraphics[width=0.32\textwidth]{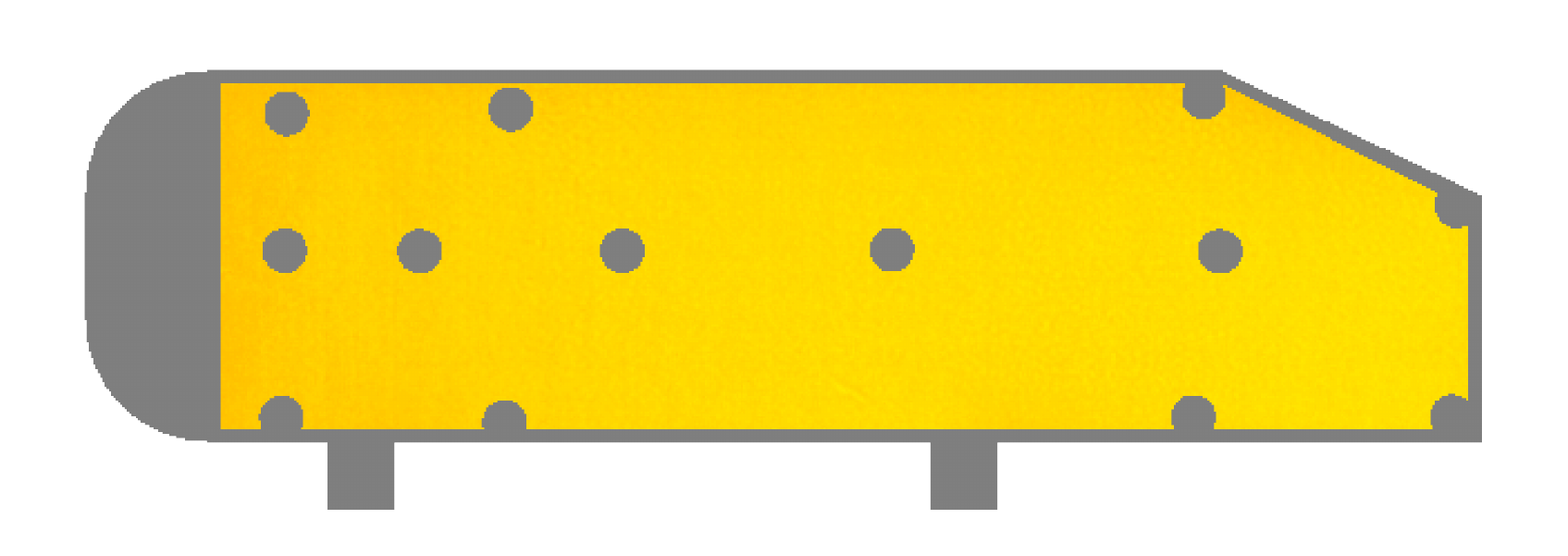}}
    \subfigure[-15 deg]{
    \includegraphics[width=0.32\textwidth]{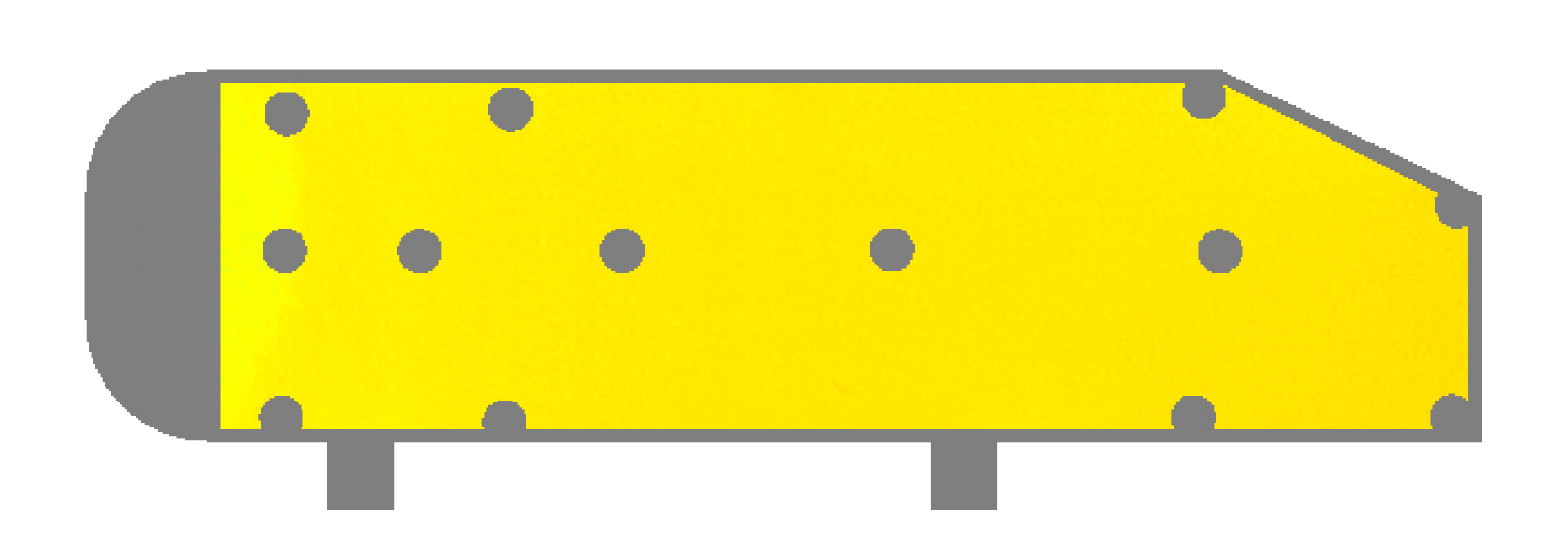}}\\
    \subfigure[-10 deg]{
    \includegraphics[width=0.32\textwidth]{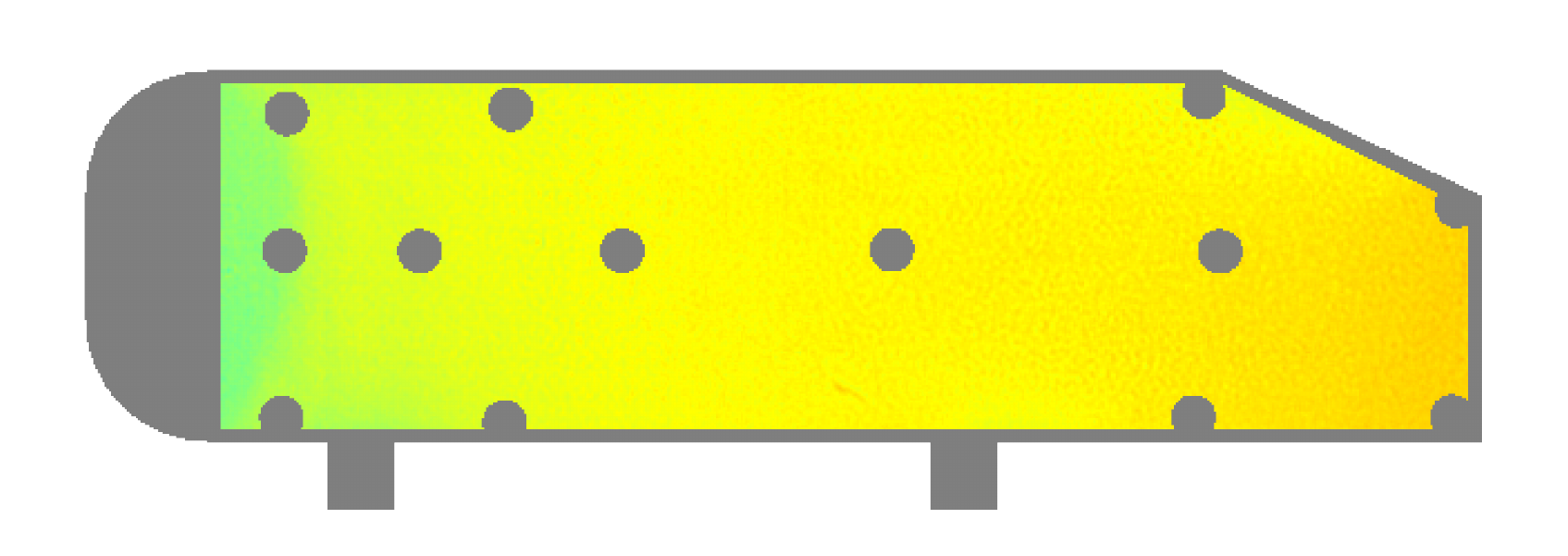}}
    \subfigure[-5 deg]{
    \includegraphics[width=0.32\textwidth]{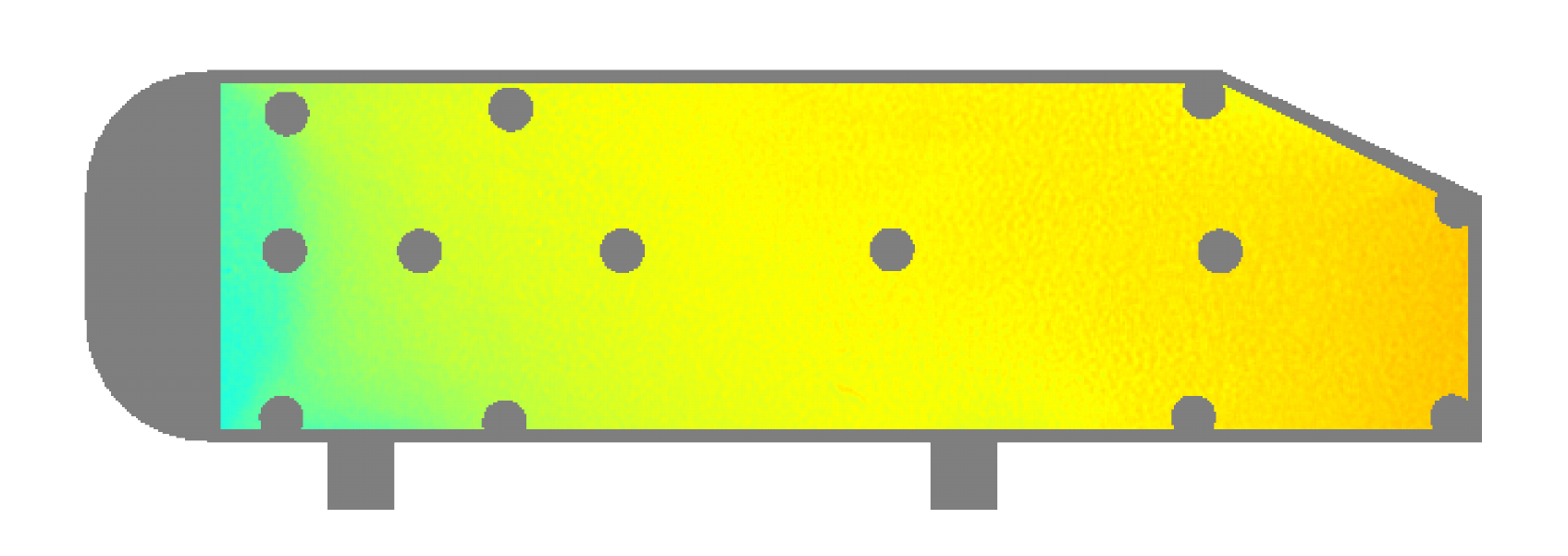}}
    \subfigure[0 deg]{
    \includegraphics[width=0.32\textwidth]{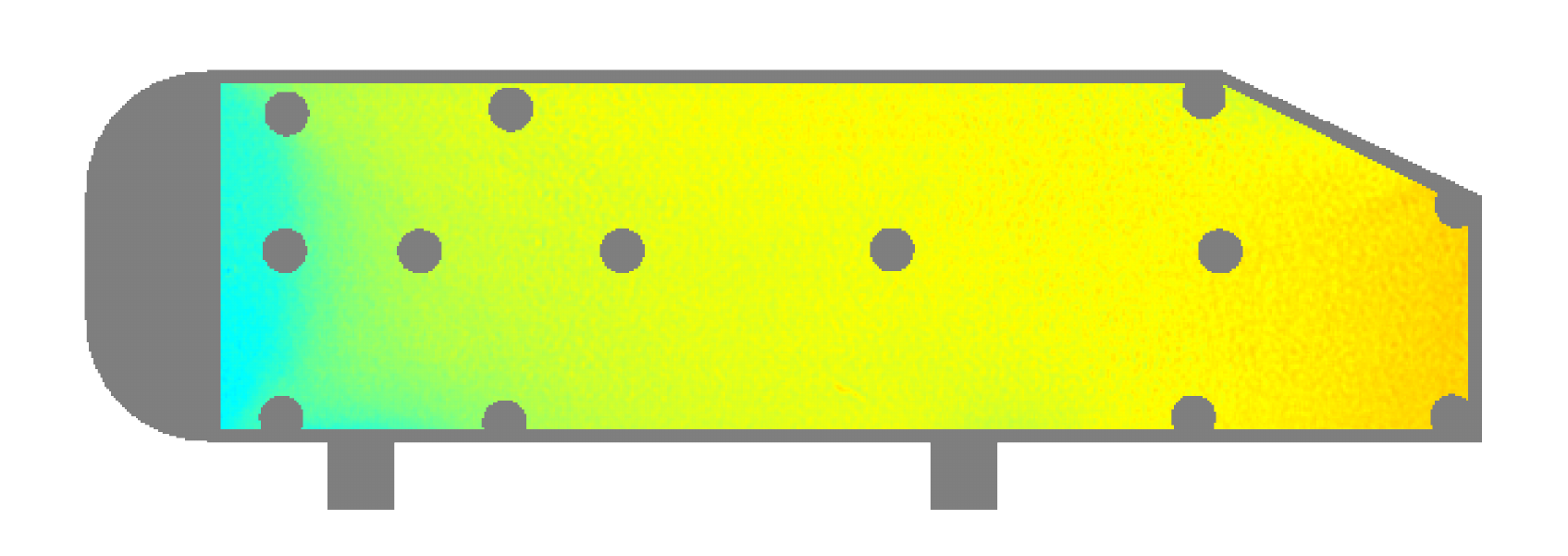}}\\
    \subfigure[5 deg]{
    \includegraphics[width=0.32\textwidth]{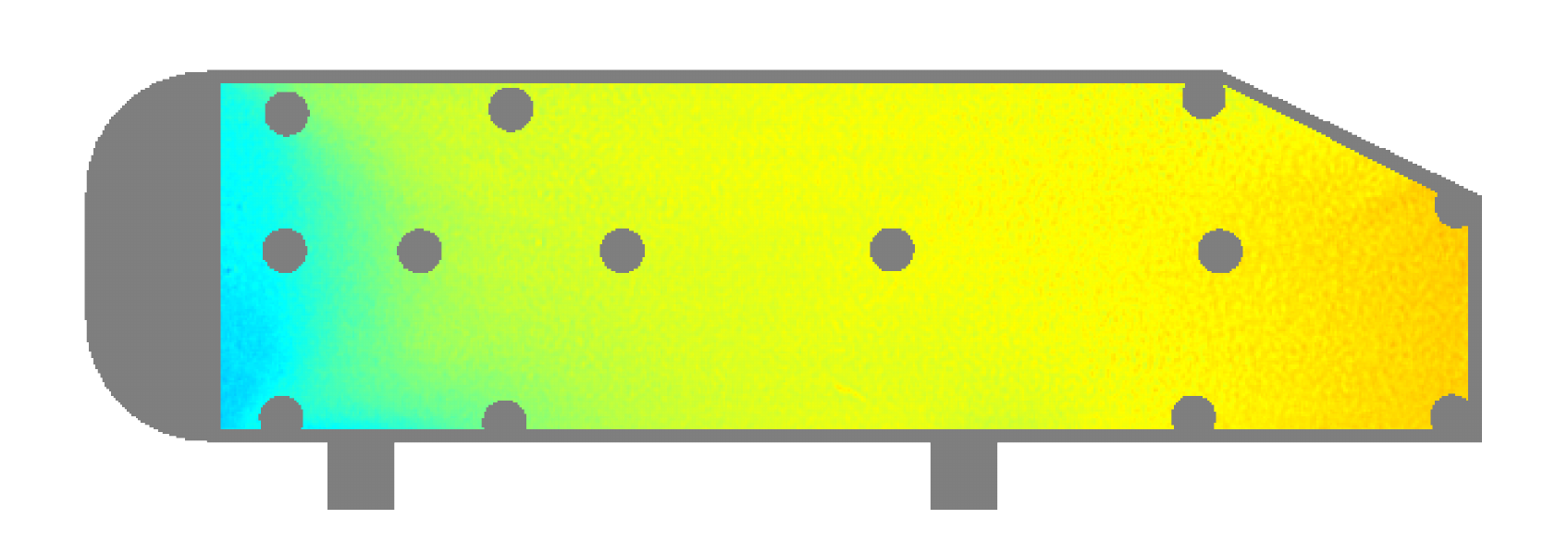}}
    \subfigure[10 deg]{
    \includegraphics[width=0.32\textwidth]{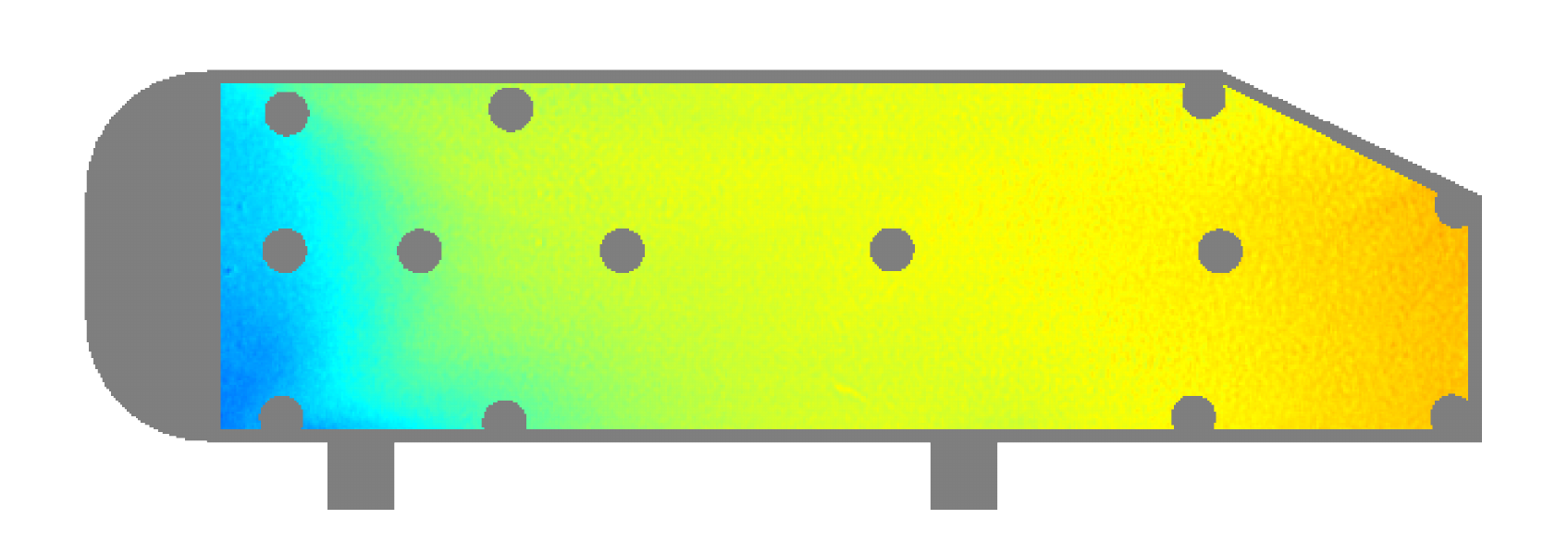}}
    \subfigure[15 deg]{
    \includegraphics[width=0.32\textwidth]{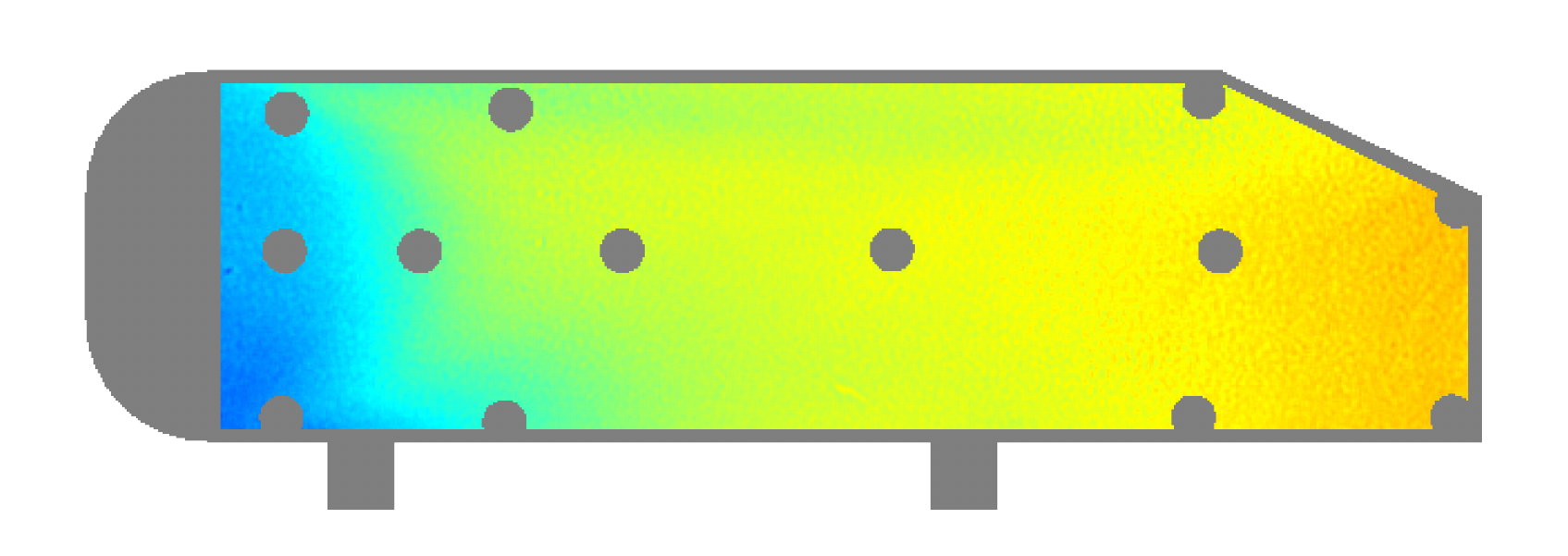}}\\
    \subfigure[20 deg]{
    \includegraphics[width=0.32\textwidth]{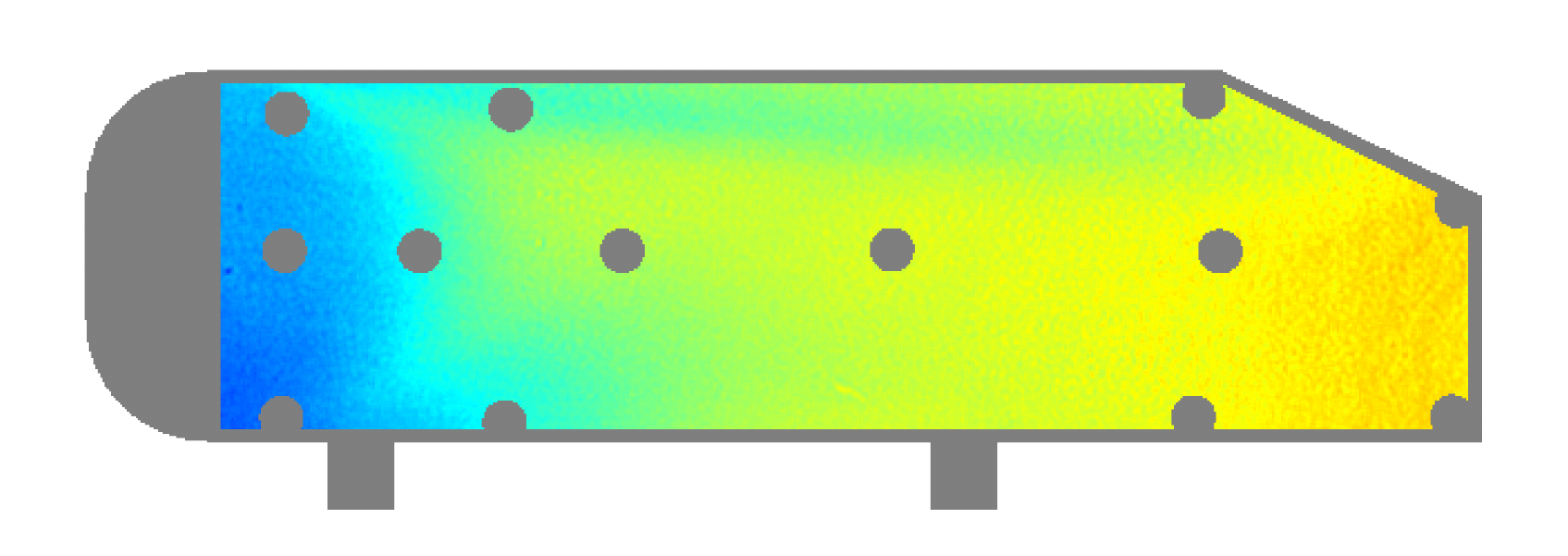}}
    \subfigure[25 deg]{
    \includegraphics[width=0.32\textwidth]{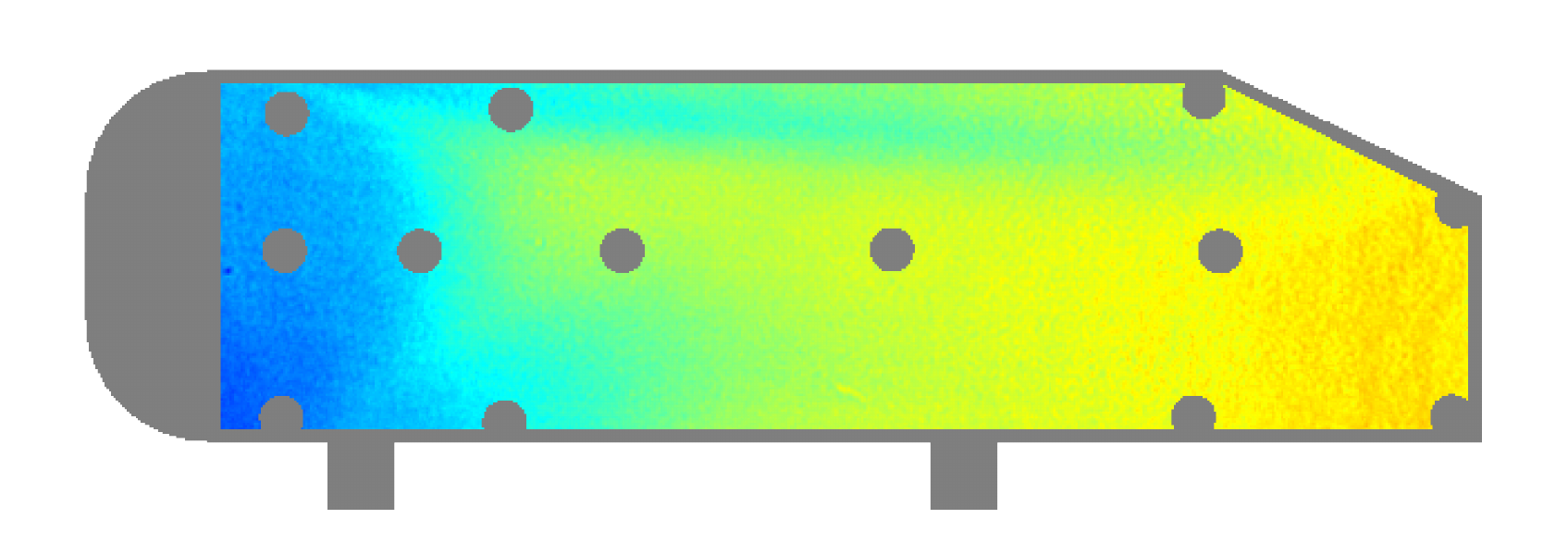}}\\
    \includegraphics[width=0.55\textwidth]{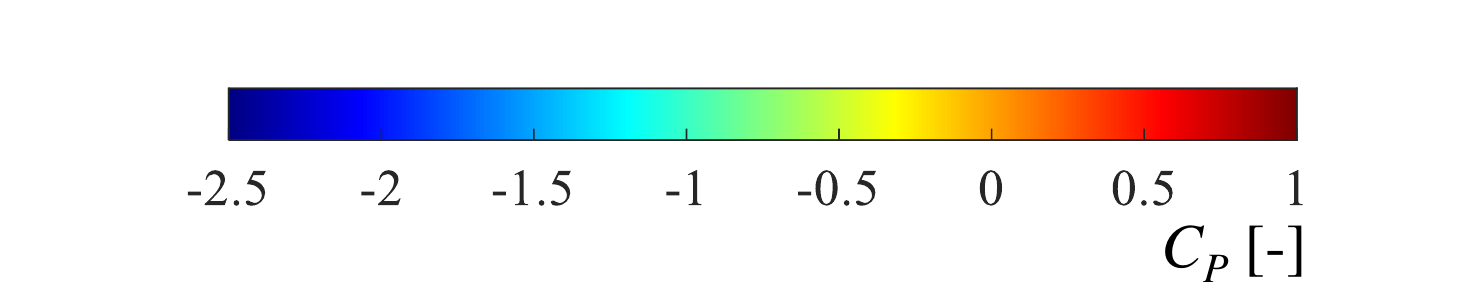}
    \caption{Pressure coefficient $C_P$ distributions on the left side of the Ahemd model from $\phi=-25$ deg to $\phi=25$ deg. Areas, where PSP is not painted, are masked in gray.}
    \label{fig:CpDis}
\end{figure}

\begin{figure}[!htbp]
    \centering
    \includegraphics[width=1.0\textwidth]{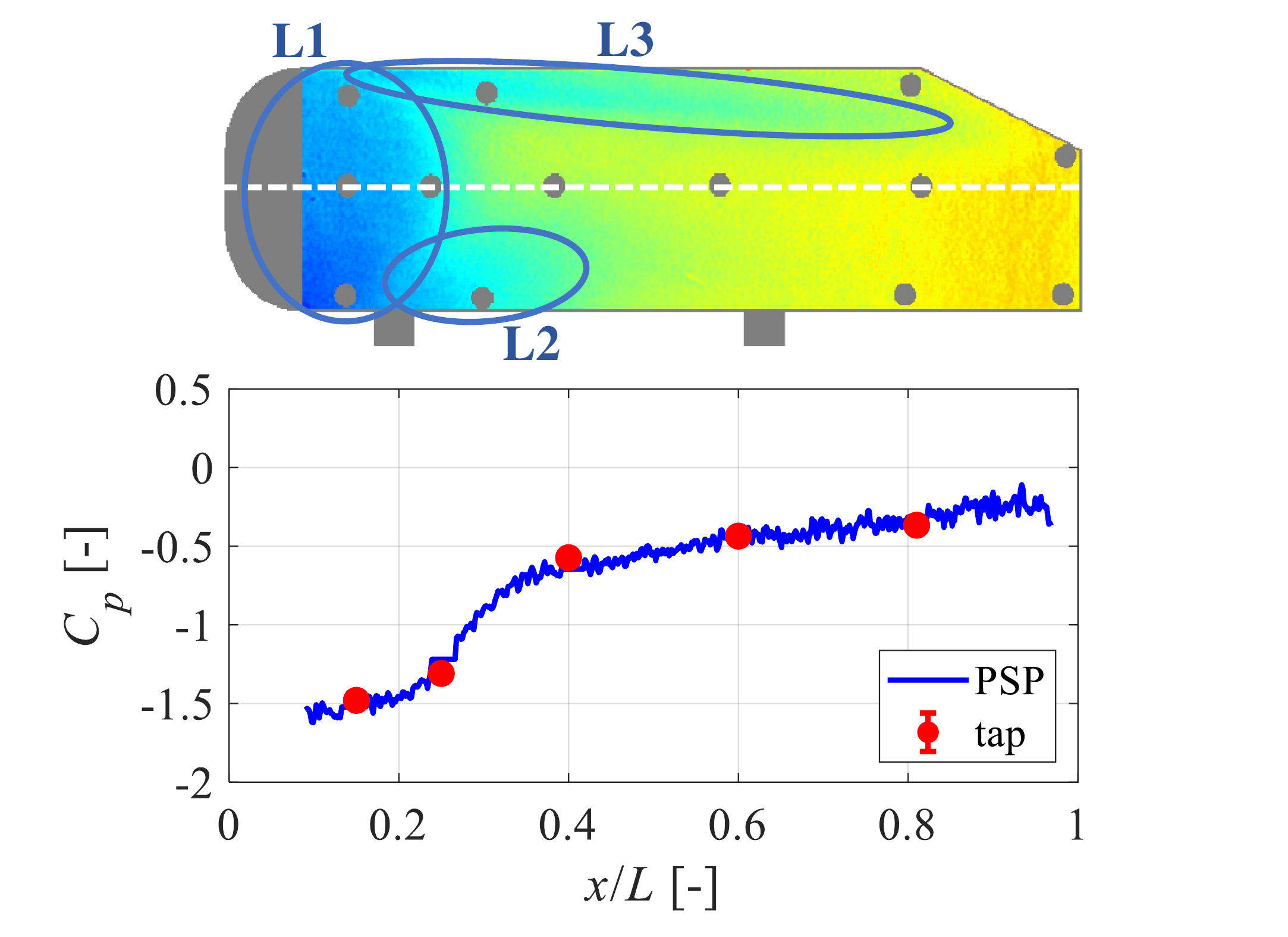}
    \caption{Structure of the $C_P$ distribution at $\phi=25$~deg (upper) and the $C_P$ profile on the white dashed line in $C_P$ distribution (lower). L1, L2, and L3 represent low-pressure regions, respectively.}
    \label{fig:CpProfile}
\end{figure}

Figure~\ref{fig:CpDis} shows the $C_P$ distributions on the left side of the Ahmed model obtained by PSP measurements. The left side of the model is the leeward and windward sides when the yaw angle is negative and positive, respectively. The areas around the image registration markers and pressure taps are masked in gray circle because the PSP was not painted. In addition, Fig.~\ref{fig:CpProfile} shows the structure of the $C_P$ distribution at $\phi=25$~deg (upper) and the $C_P$ profile on the white dashed line in $C_P$ distribution (lower).
As the yaw angle increases, the $C_P$ on the entire surface on the left side decreases, and a low-pressure region is formed at the upstream side of the model (L1) and near the front legs (L2). These low-pressure regions expand downstream as the yaw angle increases. 
In the case of $\phi=15$ deg, the low-pressure region at the top of the model (L3) becomes noticeable and becomes stronger as the yaw angle increases.
The characteristics of the $C_P$ distribution obtained in this experiment are qualitatively consistent with the results of the previous study \citep{yorita2012development} obtained under the same conditions.

The error of the PSP measurement at an arbitrary yaw angle in this experiment $\Delta C_{P,\mathrm{RMS},j}$ was calculated as shown bellow:
\begin{align}
    \Delta C_{P,\mathrm{RMS},j} =
    \sqrt{\cfrac{1}{t}
    \sum_{t=1}^\tau 
    \left(C_{P,\mathrm{PSP},t,j} - C_{P,\mathrm{tap},t,j} \right)^2
    },
    \label{eq:Cprms}
\end{align}
where, $\tau$, $C_{P,\mathrm{PSP}}$, and $C_{P,\mathrm{tap}}$ is the number of pressure taps, the pressure coefficient obtained by PSP around the pressure taps, and the pressure coefficient obtained by the pressure taps, respectively. The mean and standard deviation of $\Delta C_{P,\mathrm{RMS},j}$ were 0.059 and 0.018, respectively. Furthermore, Fig.~\ref{fig:CpProfile} shows that the shot noise in the PSP measurement data is sufficiently small, which means that these datasets can be used as training data.

\begin{figure}[!htbp]
    \centering
    \includegraphics[width=0.8\textwidth]{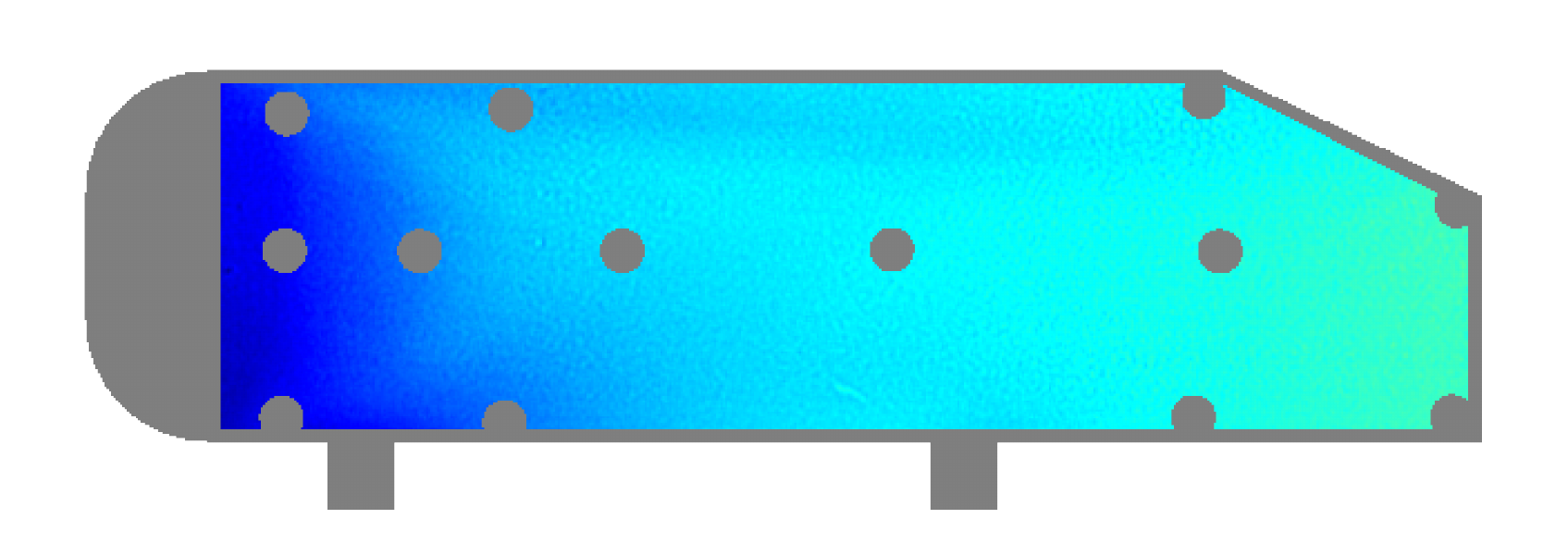}\\
    \subfigure[Mean of $\bm{X}_{\mathrm{left}}$]{
    \includegraphics[width=0.55\textwidth]{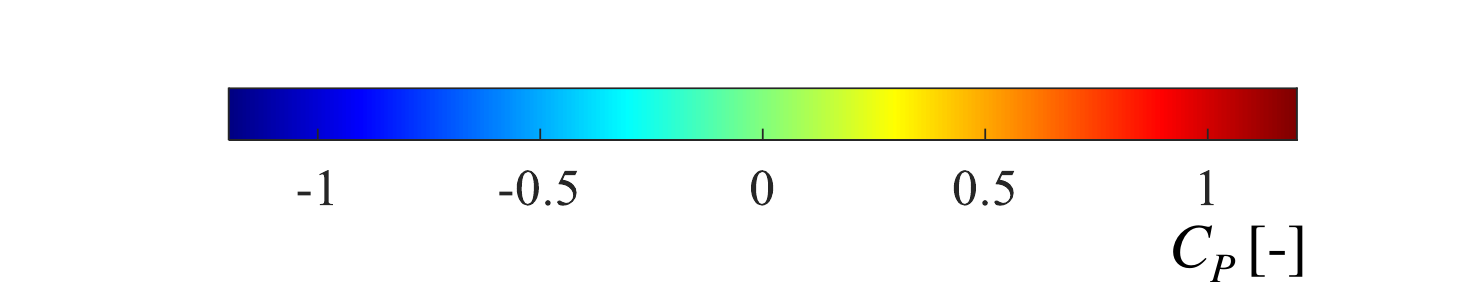}}\\
    
    \includegraphics[width=0.8\textwidth]{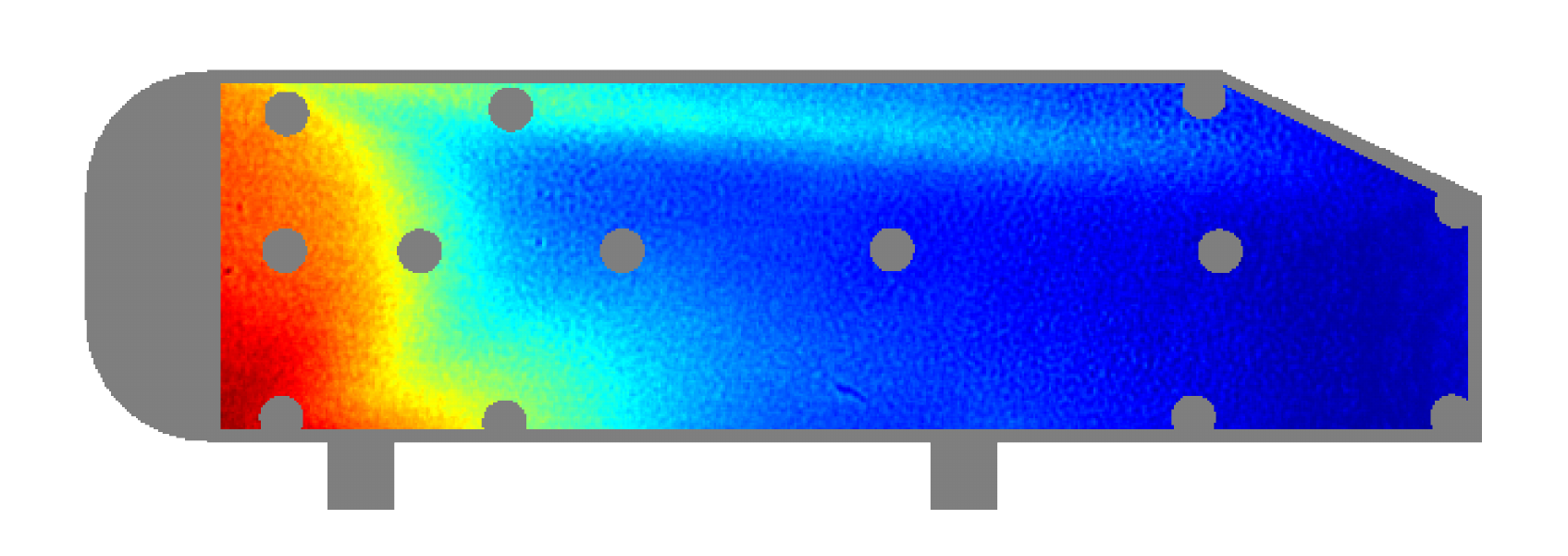}\\
    \subfigure[Standard deviation of $\bm{X}_{\mathrm{left}}$]{
    \includegraphics[width=0.55\textwidth]{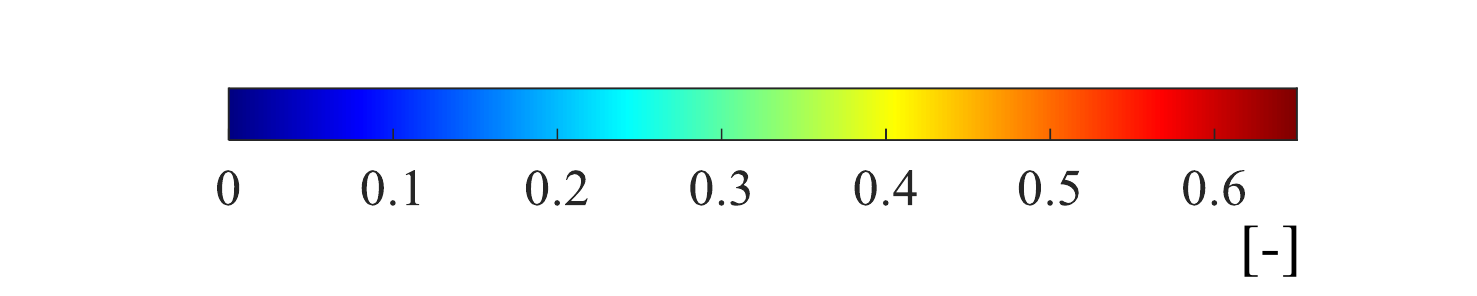}}\\
    
    \includegraphics[width=0.8\textwidth]{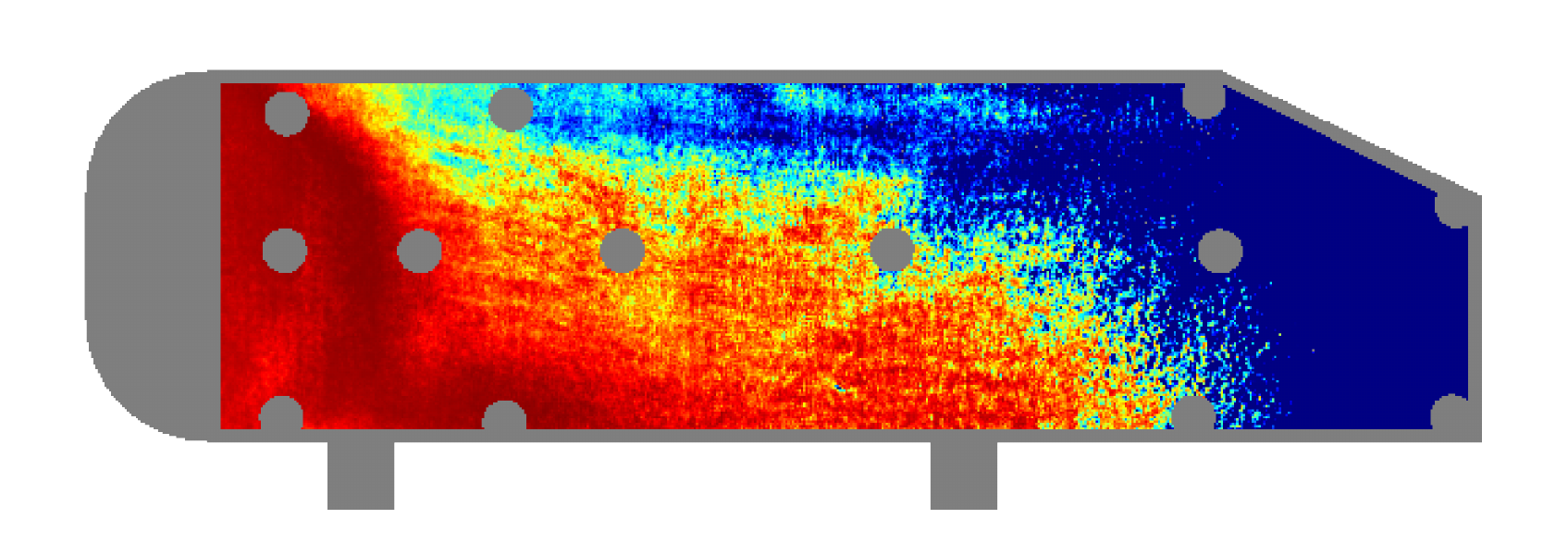}\\
    \subfigure[Coefficient of determination between $\bm{\phi}$ and $\bm{X}_{\mathrm{dif}}$]{
    \includegraphics[width=0.55\textwidth]{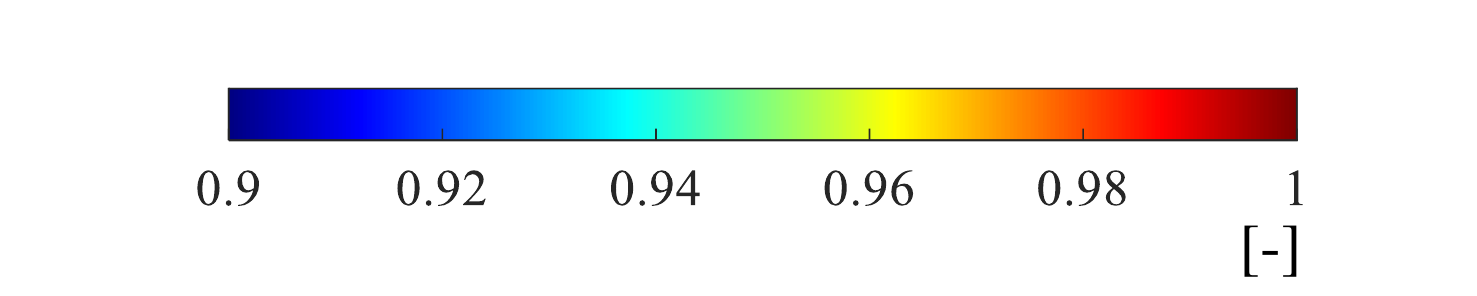}}\\
    \caption{The characteristic distributions of $C_P$ distribution}
    \label{fig:CpChrDis}
\end{figure}

Figures~\ref{fig:CpChrDis}(a)--(c) show the distributions of the mean pressure coefficient $\overline{\bm{x}}_\mathrm{both}$, the standard deviation of $\bm{X}_{\mathrm{left}}$, and the coefficient of determination between $\bm{\phi}$ and $\bm{X}_\mathrm{dif}$, respectively. The coefficient of determination, which is calculated as the square of the correlation coefficient between $\bm{\phi}$ and each row of $\bm{X}_\mathrm{dif}$, represents the strength of the linear correlation between yaw angle and the differential pressure between the left and right sides. The distribution of the standard deviation shown in Fig.~\ref{fig:CpChrDis}(b) indicates that the pressure change due to the change in the yaw angle is large in the upstream part of the model, particularly in the region near the front legs. On the other hand, the pressure change is small in the downstream part of the model. The distribution of the coefficient of determination shown in Fig.\ref{fig:CpChrDis}(c) indicates that the linear correlation between  $\bm{\phi}$ and $\bm{x}_{\mathrm{dif},i}$ is particularly strong in the upstream part of the model.

\begin{figure}[!htbp]
    \centering
    \subfigure[Mode 3]{
    \includegraphics[width=0.45\textwidth]{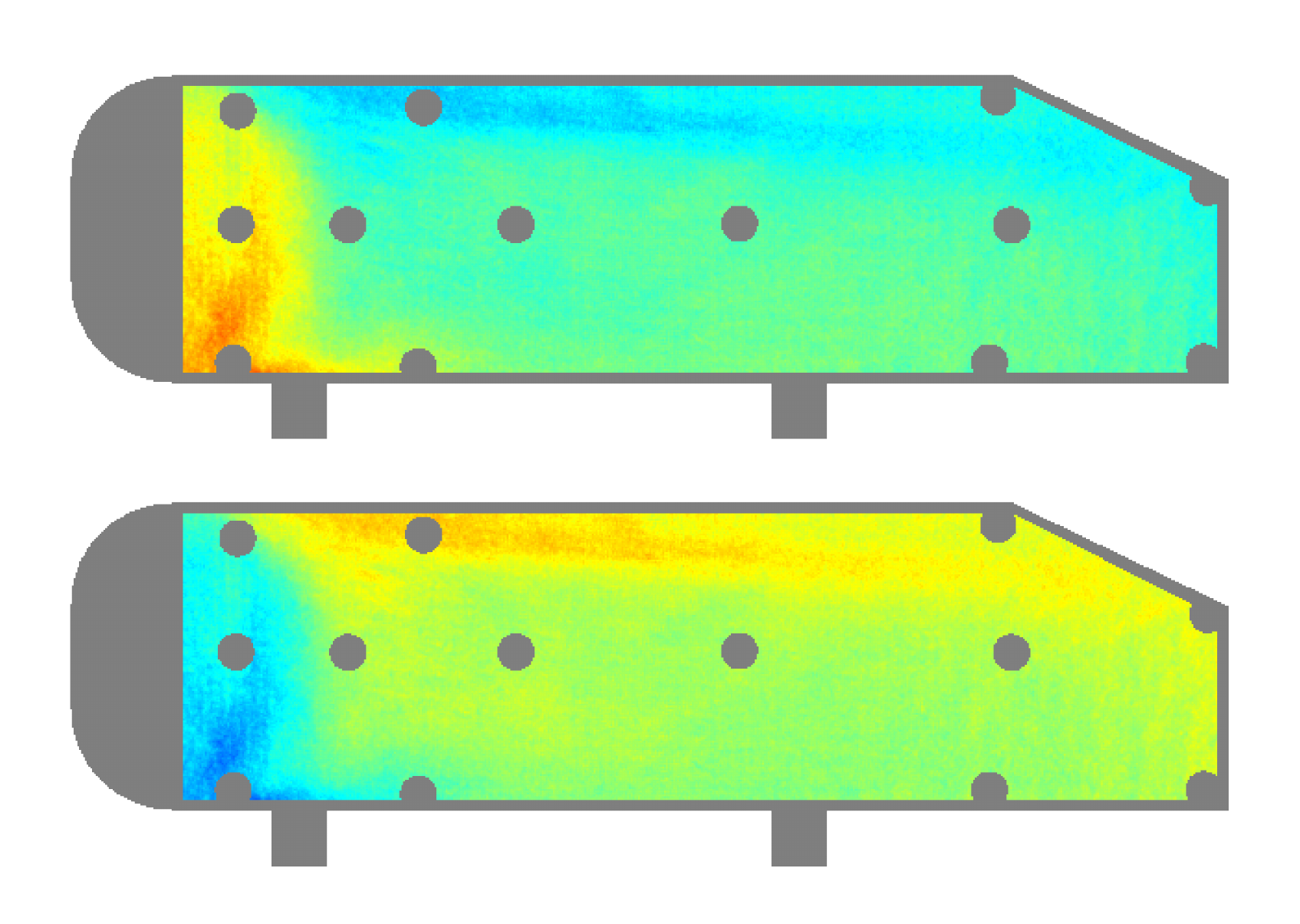}
    \includegraphics[width=0.45\textwidth]{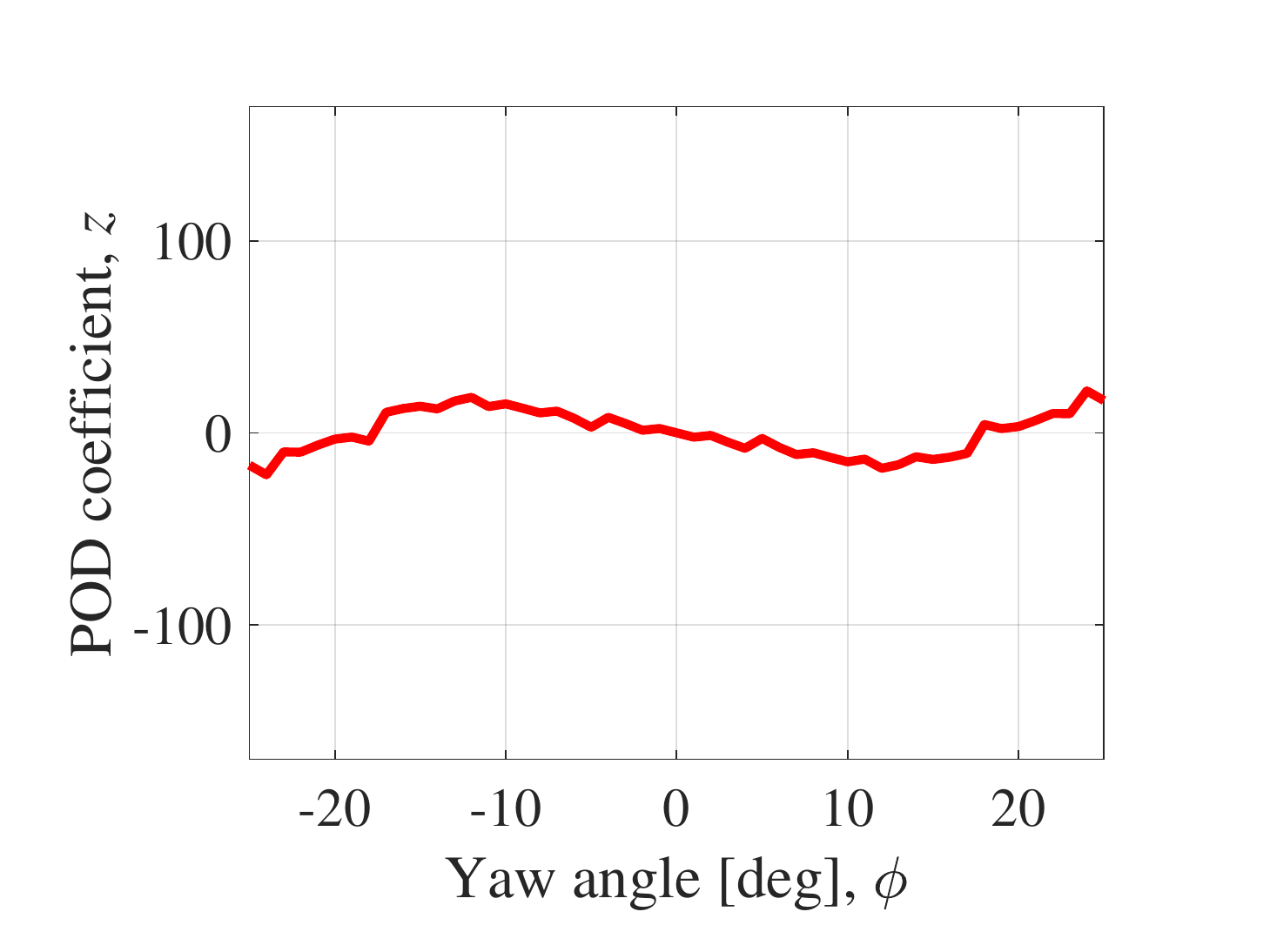}}
    \subfigure[Mode 4]{
    \includegraphics[width=0.45\textwidth]{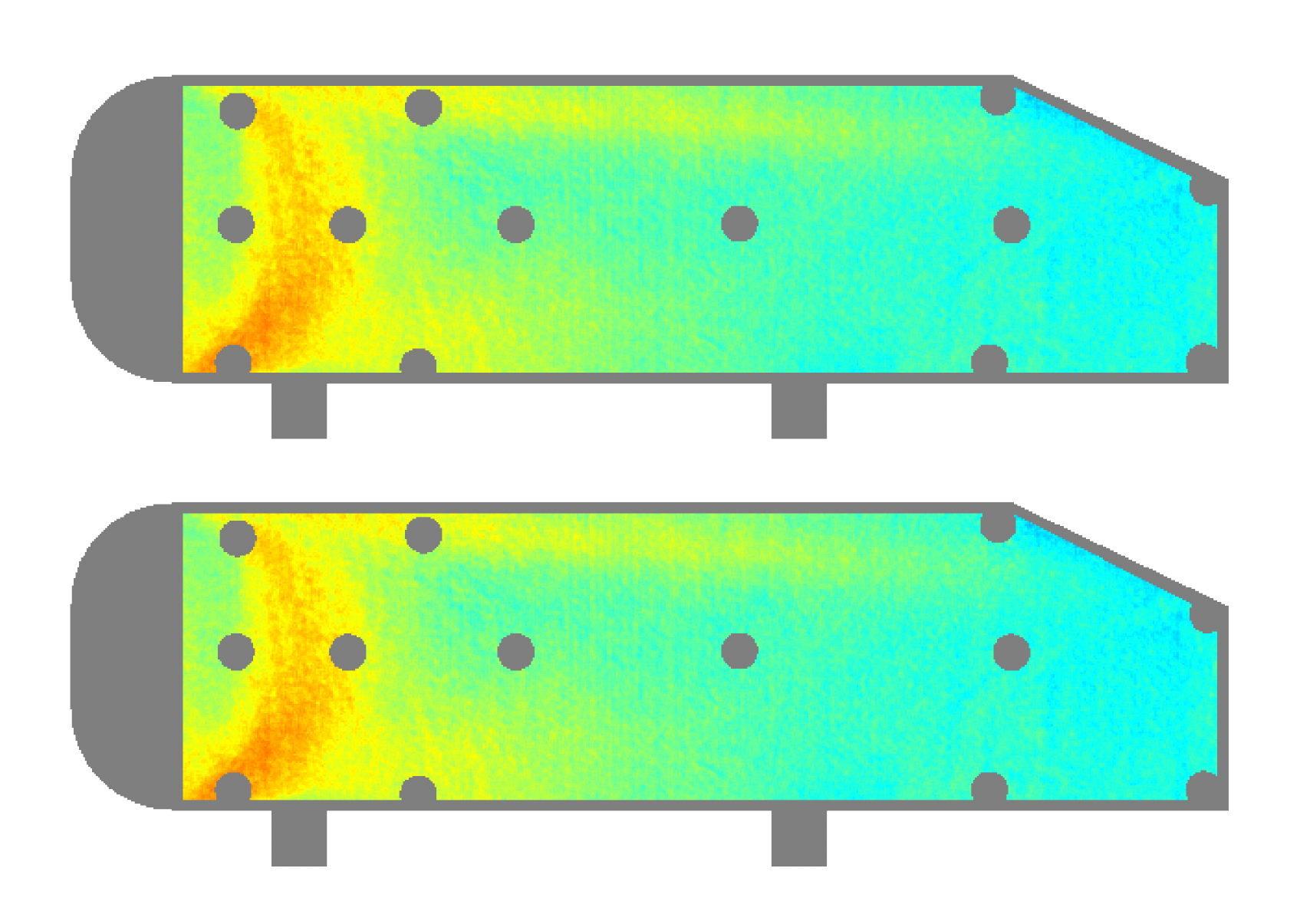}
    \includegraphics[width=0.45\textwidth]{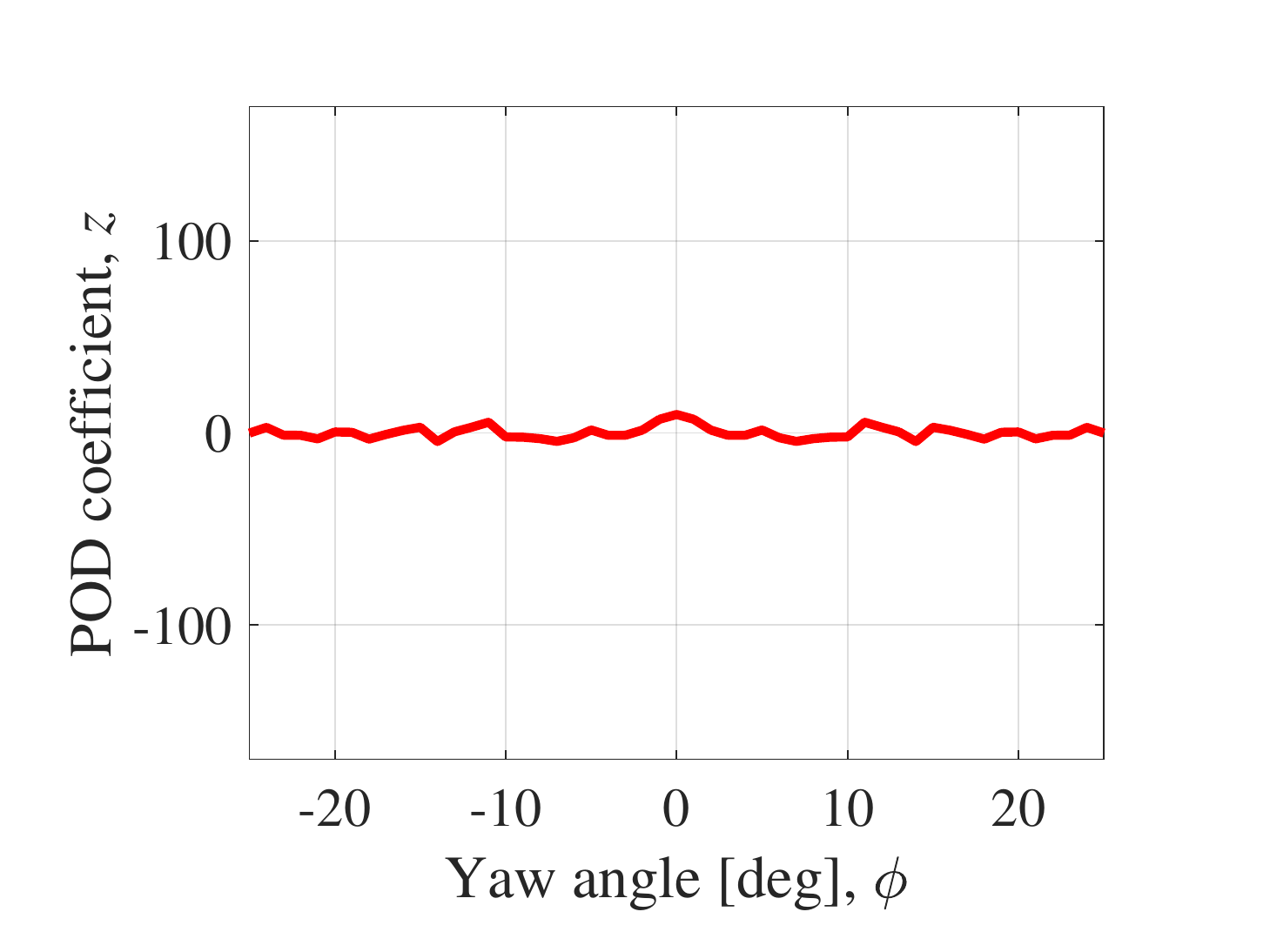}}
    \subfigure[Mode 5]{
    \includegraphics[width=0.45\textwidth]{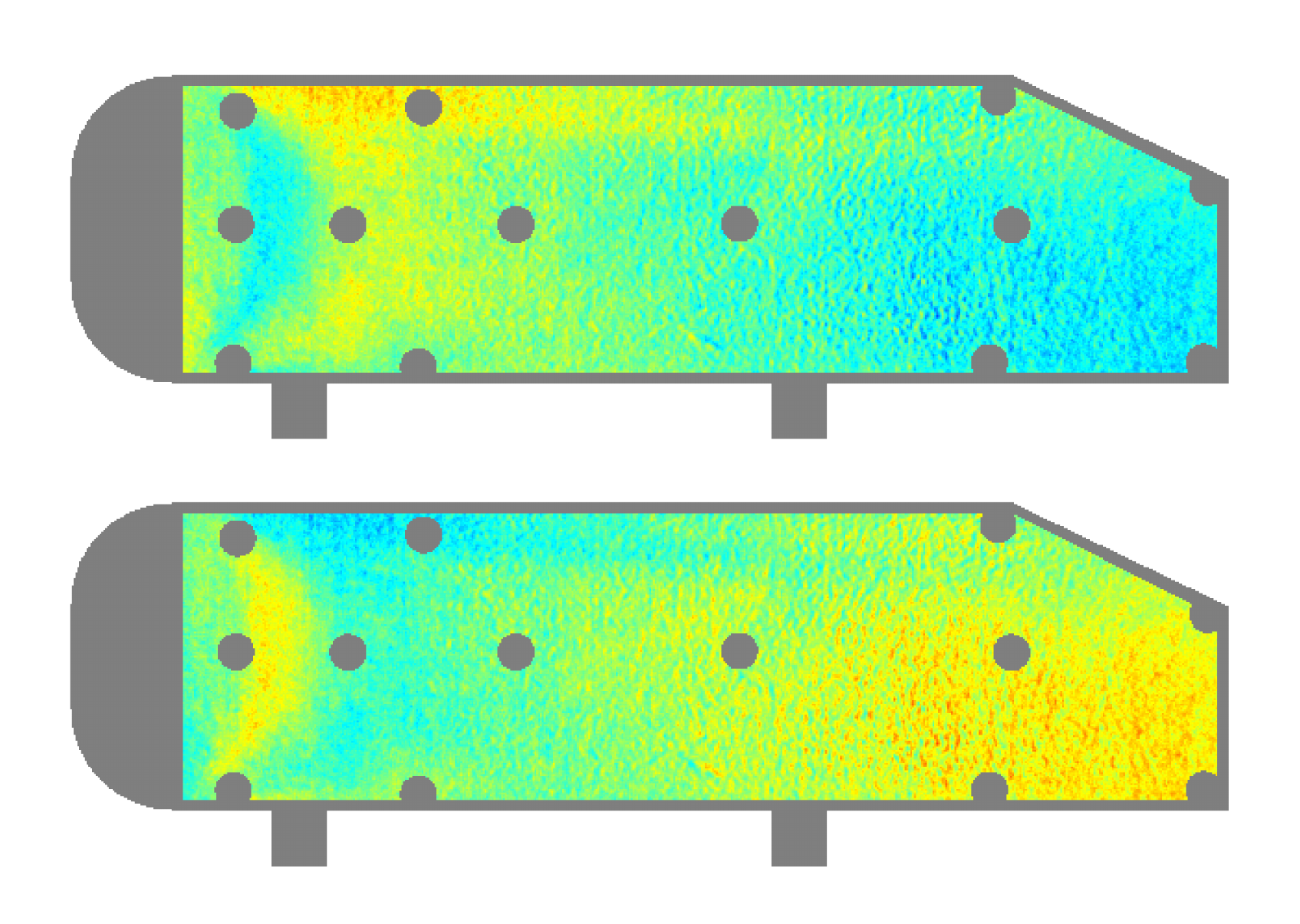}
    \includegraphics[width=0.45\textwidth]{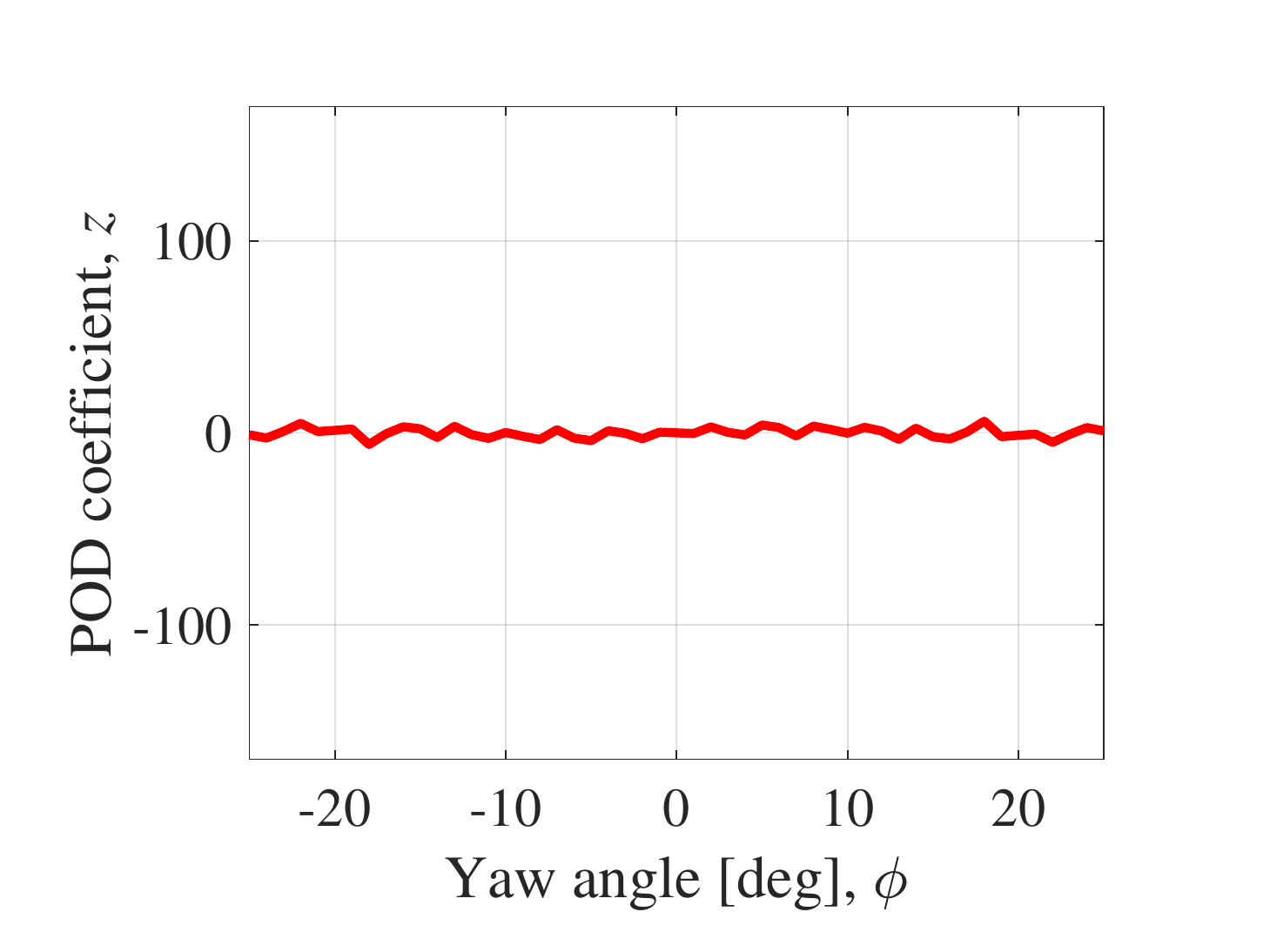}}\\
    \includegraphics[width=0.55\textwidth]{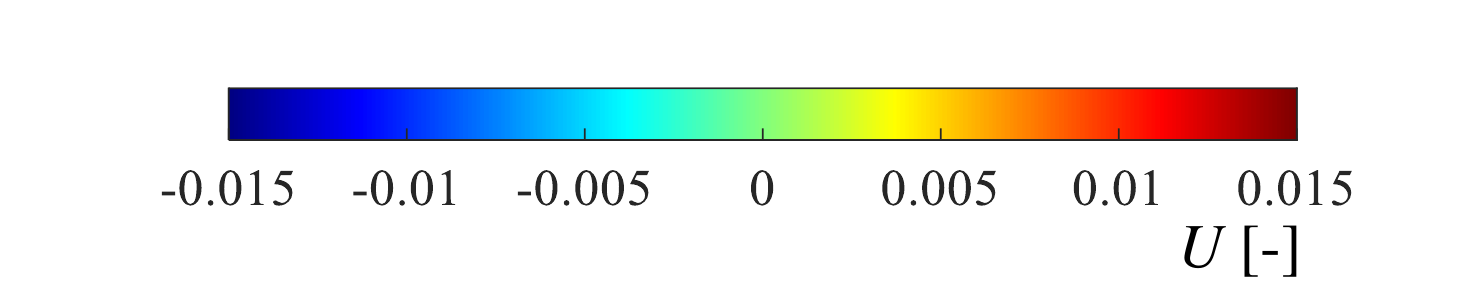}
    \caption{Results of POD : (a) contribution rate, (b)--(f) POD spatial mode distribution of left side (upper left), one of right side (lower left) and POD coefficient (right)}
    \label{fig:POD}
\end{figure}

Figure~\ref{fig:POD} shows the POD modes of $\widetilde{\bm{X}}_{\mathrm{both}}$ calculated by Eq.~\eqref{eq:Xboth2}. The odd-numbered spatial POD modes have positive and negative reversals on the left and right sides, while the even-numbered spatial POD modes are identical on the left and right sides. Therefore, the POD coefficients for odd- and even-number modes are antisymmetric and symmetric regarding the yaw angle, respectively. Subfigure~(a) shows the contribution rate of the first mode is dominant at 95.7\%, indicating that the POD efficiency of this data is high. In addition, subfigure~(b) shows a strong correlation between the first mode coefficient and the yaw angle. Therefore, most of the changes in the pressure field due to changes in yaw angle are represented by the first mode. These mean that the low-dimensional model works well for the sparse sensor placement and the estimations. The sum of the contribution rates of the first to fourth modes is 99.7\%, and the number of modes $r$ for the low-dimensional model of the $C_P$ distribution was set to four in the following discussion.

\subsection{Selected sensor pair positions and reconstructed pressure distribution}
\begin{figure}[H]
    \centering
    \subfigure[The random selection method]{
    \includegraphics[width=0.95\textwidth]{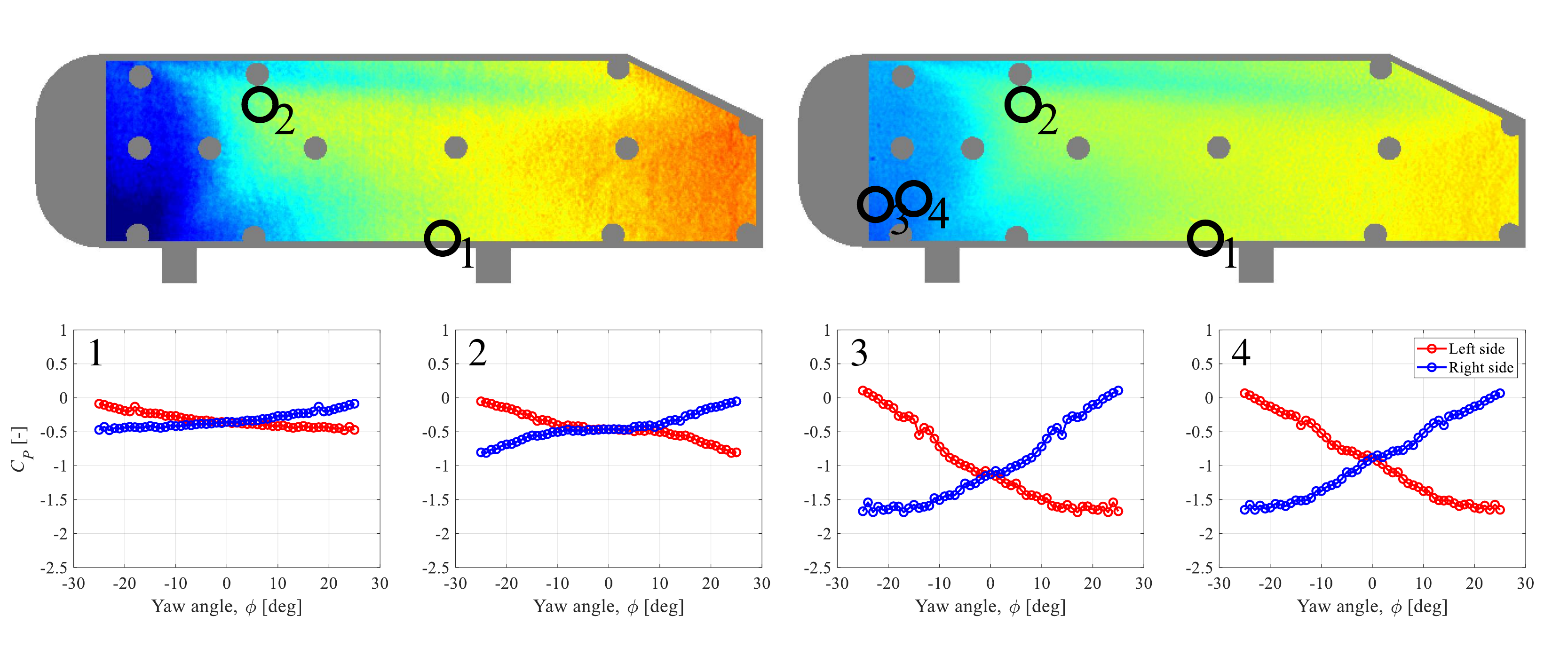}}\\
    \includegraphics[width=0.55\textwidth]{fig_X_ColorBar.pdf}
    \caption{Sensor pair positions selected by each algorithm and $C_P$ distribution at $\phi=25$~deg reconstructed by two (upper left) or four (upper right) selected sensor pairs in the case of $r=4$ and $q=1$. The order of selected sensor pairs are displayed with the sensors. In addition, the lower figures show the input of $C_P$ at each sensor pair location.}
    \label{fig:CensLoc}
\end{figure}

The upper figures in Fig.~\ref{fig:CensLoc} show the sensor positions selected by the OMP algorithm, the DG-vector algorithm, the hybrid algorithm, and the random selection method (Random), and the $C_P$ distributions at $\phi=25$~deg reconstructed by two (upper left) or four (upper right) selected sensor pairs in the case of $r=4$. It should be noted that the other sides are not shown, and therefore, the two and four sensor pair corresponds to four and eight symmetric sensors in both sides, respectively. The selected order of sensor pairs is displayed with the sensors. In addition, the lower figures in Fig.~\ref{fig:CensLoc} show the input of $C_P$ at each sensor point and each yaw angle. Here, the result of the random selection method is one trial example and shown as the reference.

The OMP algorithm selected the first sensor pair position in which the pressure change has the strongest linear correlation with the yaw angle. 
The coefficient of determination at the first sensor pair point was 0.999.
On the other hand, the OMP algorithm selected the point with the smaller pressure change after the second sensor pair position. This is because the OMP algorithm minimizes the residual of the estimated yaw angle at each step, and the linear correlation between the yaw angle and the pressure is strong at the first sensor pair position. The second and subsequent sensor pairs are fitted to the noise component of the pressure data, thus points with a small pressure change due to the yaw angle change were selected.

The DG-vector algorithm selected the first sensor pair at the lower part of the upstream side where the pressure change was large, and the second sensor pair was selected at the base of low-pressure region L3. The third and fourth sensor pairs were selected near the first and second sensor pairs. For all sensor pairs, the point with the large pressure change was selected.

Since the pressure changes of the second and subsequent sensor pairs selected by the OMP algorithm were low, the hybrid algorithm was set to $q=1$, and the second and subsequent sensor pairs were determined by the DG-vector algorithm. Therefore, the first sensor pair selected by the hybrid algorithm is the same as that selected by the OMP algorithm, and the second sensor pair selected by the hybrid algorithm is different from that selected by the OMP algorithm. The third and subsequent sensor pairs selected by the hybrid algorithm were chosen from the vicinity of sensor pair positions obtained by the DG-vector algorithm. For all sensor pairs, the points with the large pressure change were selected.

The $C_P$ distributions reconstructed by the DG-vector and hybrid algorithms ($p=2,4$) are almost the same as the original $C_P$ distribution in Fig,~\ref{fig:CpDis}(k). On the other hand, the $C_P$ distribution reconstructed by the random selection method ($p=2$) was overestimated compared to the original $C_P$ distribution. These results show that sensor position optimization is effective for pressure distribution reconstruction.

\subsection{Dependence of estimation error on the number of sensor pairs}
\begin{figure}[!htbp]
    \centering
    \subfigure[the validation I (w/o leave-one-out cross-validation)]{
    \includegraphics[width=0.8\textwidth]{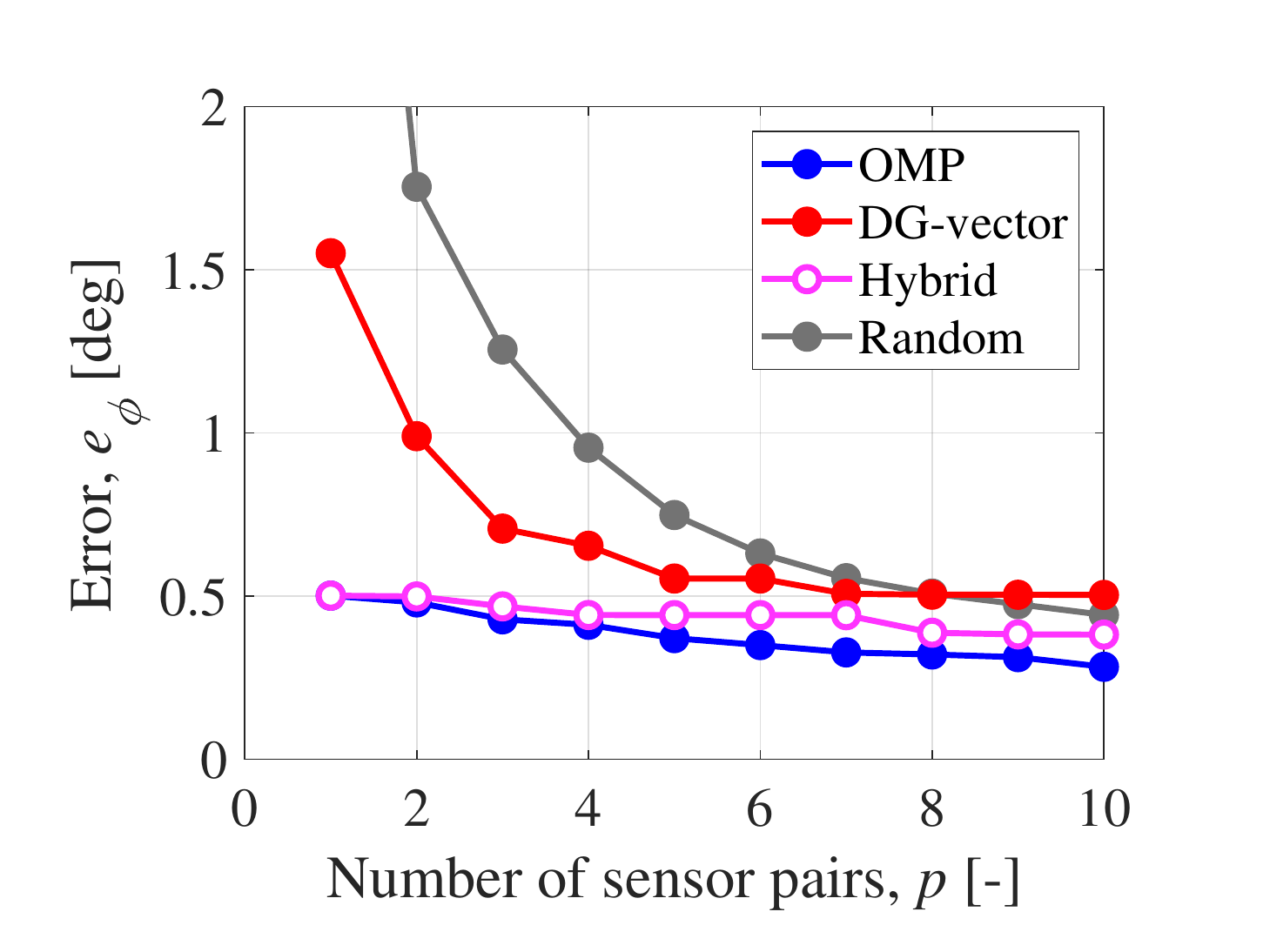}}\\
    \subfigure[the validation II (w/ leave-one-out cross-validation)]{
    \includegraphics[width=0.8\textwidth]{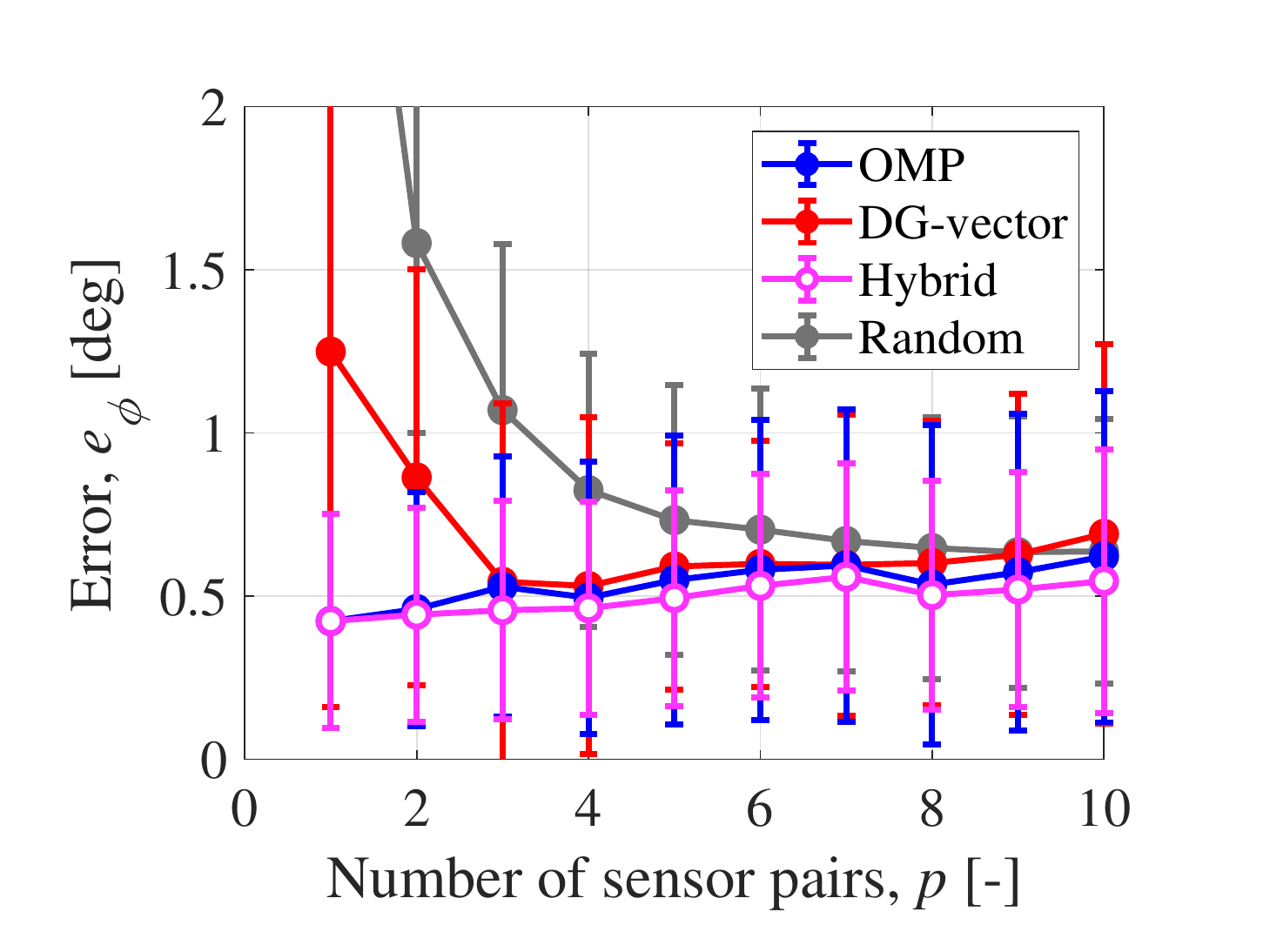}}
    \caption{Estimation error of yaw angle against number of sensor pairs}
    \label{fig:YawEstError}
\end{figure}

\begin{figure}[!htbp]
    \centering
    \subfigure[the validation I (w/o leave-one-out cross-validation)]{
    \includegraphics[width=0.8\textwidth]{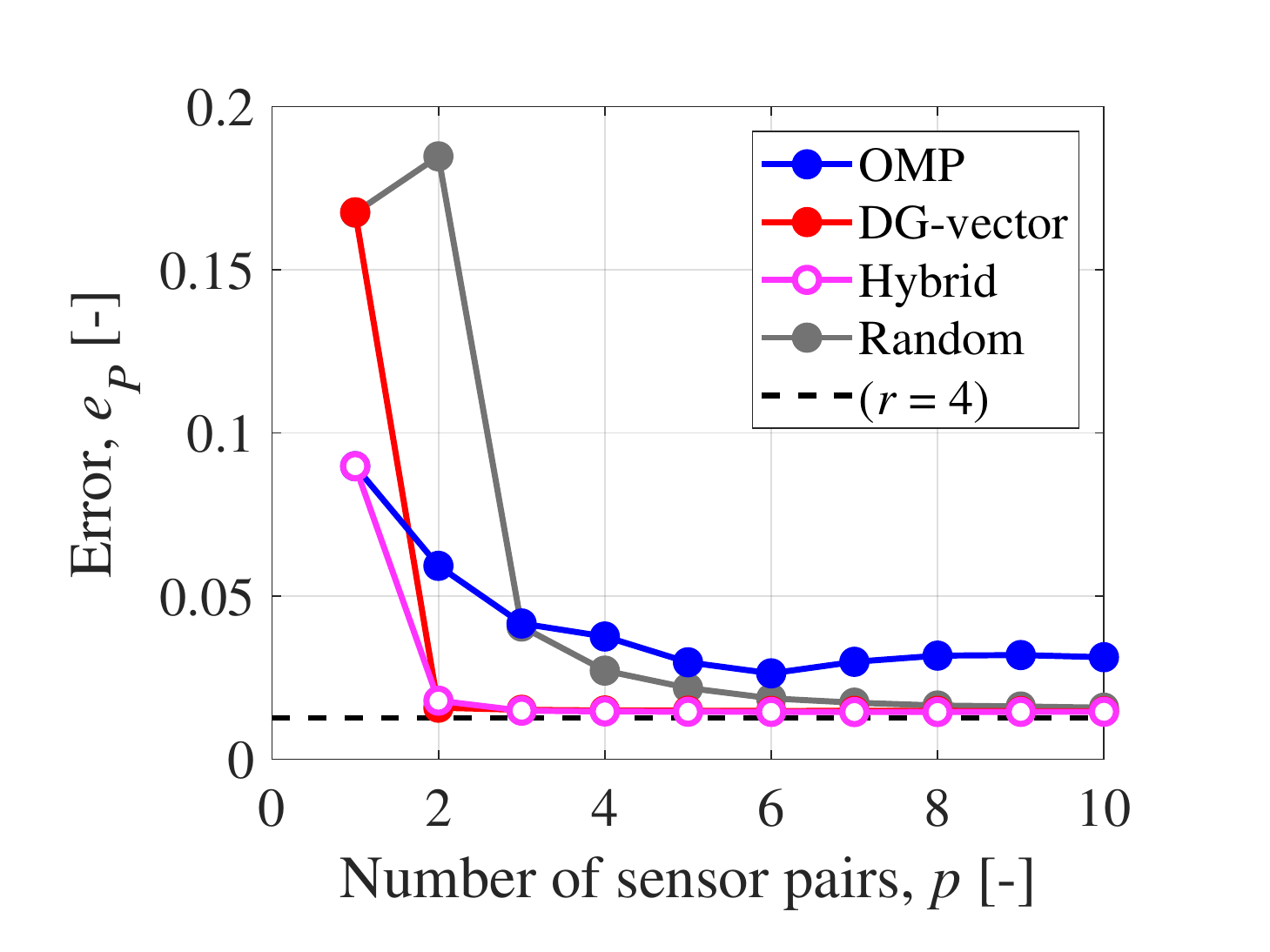}}\\
    \subfigure[the validation II (w/ leave-one-out cross-validation)]{
    \includegraphics[width=0.8\textwidth]{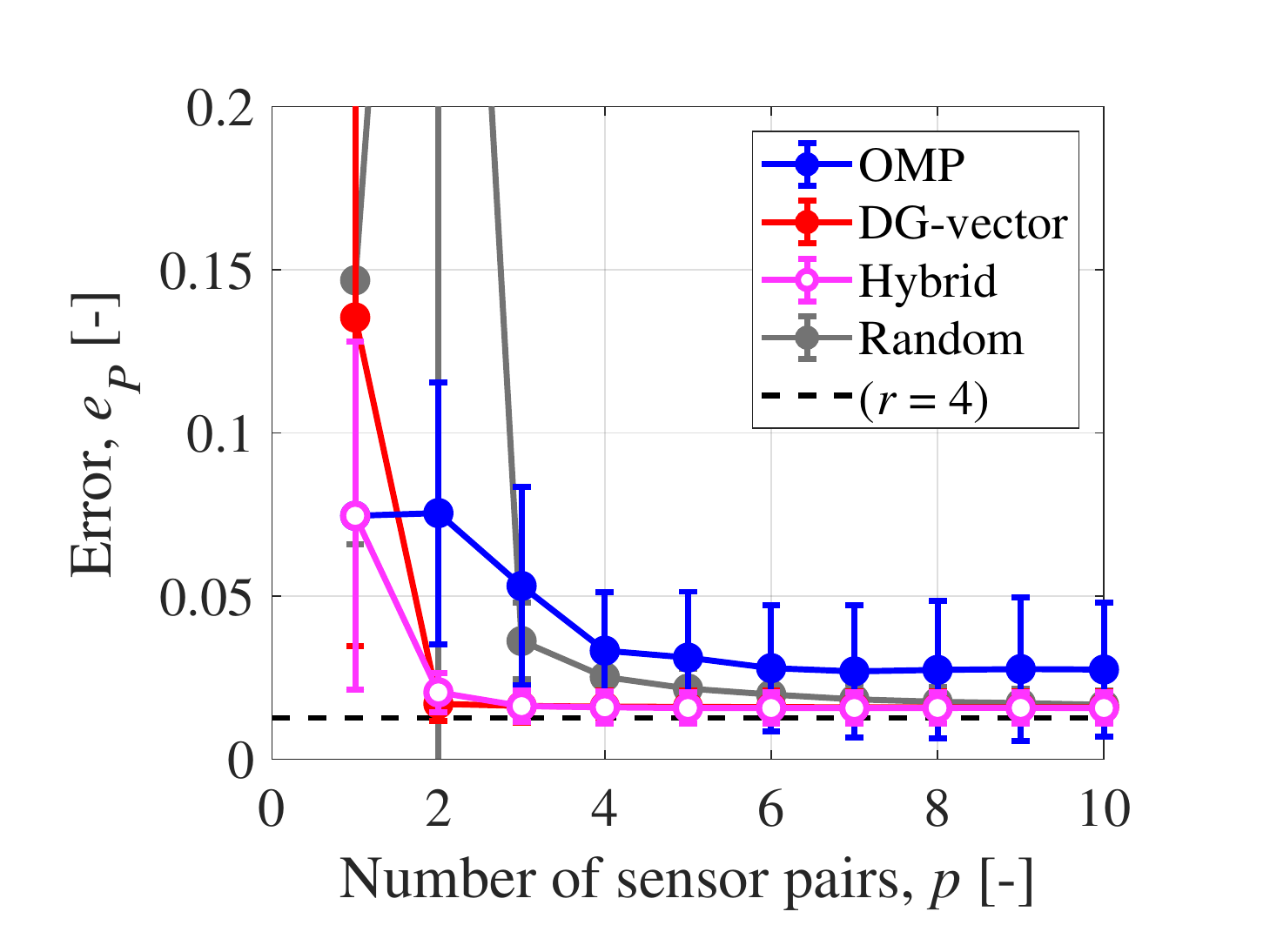}}
    \caption{Reconstruction error of $C_P$ distribution against number of sensor pairs}
    \label{fig:CpEstError}
\end{figure}

Figures~\ref{fig:YawEstError} and \ref{fig:CpEstError} show the relationship between the number of sensor pairs and the errors of the yaw angle estimation and pressure distribution reconstruction, respectively. Here, subfigures (a) and (b) correspond to the results of the validation I and II, respectively. In the validation II, the error bars represent the mean and standard deviation of the cross-validation. Here, $r=4$ was set for the pressure distribution reconstruction, and $q=1$ was set for the hybrid algorithm. 
The estimation error of the random selection method is averaged over 100 trials.
The dashed line in Fig.~\ref{fig:CpEstError} represents the error when observation is possible at all points is shown, and it corresponds to the error caused by constructing the low-dimensional model with four POD modes.

Figure~\ref{fig:YawEstError} illustrates that the estimation error of the yaw angles decreases as the number of sensor pairs increases for all algorithms in the validation I, but this is not the case in the validation II. 
The algorithms that minimize the error for two or more sensor pairs are the OMP algorithm for the validation I and the hybrid algorithm for the validation II, respectively. The increase in error due to the increase in the number of sensor pairs is caused by overfitting in validation II. As a result of fitting the estimation model to the noise component in the training data, the error increases for unknown data. In the OMP algorithm, the effect of overfitting is especially pronounced because the pressure changes of the second and subsequent sensor pairs are low. The hybrid algorithm, on the other hand, still selects sensitive points after the second sensor pair, and thus, the error of the hybrid algorithm in the validation II is expected to be smaller than that of the OMP algorithm. In terms of the yaw angle estimation, the minimum error is achieved with the single sensor pair selected by the OMP and hybrid algorithms, and the use of two or more sensor pairs hardly decreases the error. 

Figure~\ref{fig:CpEstError}, which shows the reconstruction error of the $C_P$ distribution, illustrates that the errors of the DG-vector and hybrid algorithms are lower than the PSP measurement error of 0.054 at the number of sensor pairs of two. The range of error bars is small, and the pressure distribution reconstruction is stable. In addition, the error hardly decreases even if the number of sensor pairs is three or more. On the other hand, when the number of sensor pairs is three, the reconstruction error obtained by the OMP algorithm is larger than that of randomly selected sensor pairs. Since the range of the error bar is also relatively large, the OMP algorithm is not suitable for pressure distribution reconstruction. The cases of two sensor pairs selected by the DG-vector and hybrid algorithms are optimal from the viewpoint of the pressure distribution reconstruction. 

The results and the discussions above show that the case of the hybrid algorithm with two sensor pairs is most suitable for yaw angle estimation and/or pressure distribution reconstruction.

\section{Conclusions}
In this study, sparse sensor position optimization for yaw angle estimation and pressure distribution reconstruction was performed using the time-averaged pressure coefficient distribution data at various yaw angles on the Ahmed model surface acquired by the PSP measurements at various yaw angles. The linear regression for the yaw angle and the least-squares estimation of the POD mode coefficients, based on data-driven sparse sampling, were conducted. The pressure distributions were reconstructed by the estimated POD mode coefficients together with the spatial POD modes. Three different sensor position optimization algorithms, the OMP algorithm, the DG-vector algorithm, and the hybrid algorithm, which combines the previous two algorithms, were compared and evaluated. The objective functions of OMP and DG were the indices regarding the yaw angle estimation and the POD mode coefficient estimation, respectively. It should be noted that the symmetric sensors on the left and right sides of the model were assumed. 

The selected sensor pair positions showed the characteristics of each algorithm. As a result of evaluating the generalization performance in terms of the yaw angle estimation, the OMP and hybrid algorithms with one sensor pair provided the smallest error, though there is no advantage in increasing the number of sensor pairs to two or more in the problem setting of this study due to overfitting. In terms of the pressure distribution reconstruction, the DG-vector and hybrid algorithms with two sensor pairs are effective. These results indicate that the hybrid algorithm with two sensor pairs is the most effective for the yaw angle estimation and/or the pressure distribution reconstruction. This suggests the effectiveness of combining multiple greedy algorithms for different objectives. Synergistic effects can be expected by understanding the characteristics of each algorithm and combining them appropriately.

The technique proposed in this study can be applied not only to the Ahmed model but also to other vehicle body shapes. Once a pressure database is obtained either by an experiment or numerical analysis as in this study, sensor pair positions that effectively estimate the yaw angle and the reconstructed pressure distribution can be systematically obtained.
Furthermore, in addition to the yaw angle estimation and the pressure distribution reconstruction, other physical quantities such as flow velocity can be estimated by applying this technology. However, it should be noted that we proposed a data-driven method in this study, and it is not expected to work properly under the conditions that the input is extrapolated from training data, for example, when the yaw angle of the wind direction is 30 degrees, when there is a pitch angle, or when the wind speed is significantly different. These will be the subject of challenging future research.

\section*{Acknowledgment}
T.~Nonomura and Y.~Iwasaki were supported by Japan Science and Technology Agency, FOREST Grant Number, JPMJFR202C. 
Y.~Saito was supported by Japan Science and Technology Agency, ACT-X Grant Number, JPMJAX20AD.
T.~Nagata was supported by Japan Science and Technology Agency, CREST Grant Number JPMJCR1763. 
Y.~Iwasaki was supported by Japan Society for the Promotion of Science, KAKENHI Grant No. JP21J21207.
Y.~Ozawa was supported by Japan Society for the Promotion of Science, KAKENHI Grant No. JP20H0278.

\bibliography{main}

\clearpage

\clearpage

\clearpage

\clearpage

\clearpage

\clearpage

\clearpage
\begin{figure}[!htbp]
    \centering
    \subfigure[Contribution rate]{
    \includegraphics[width=0.6\textwidth]{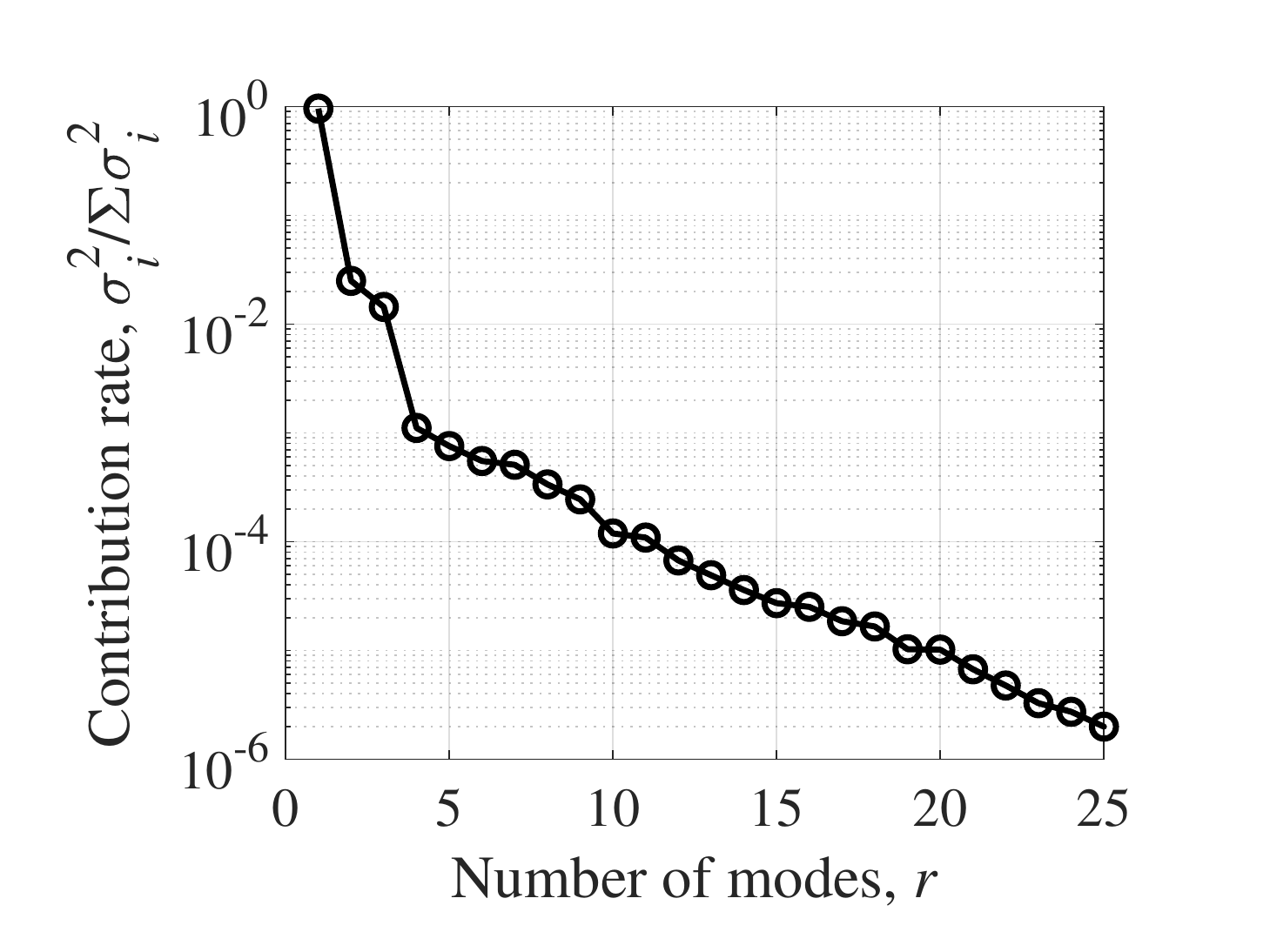}}
    \subfigure[Mode 1]{
    \includegraphics[width=0.45\textwidth]{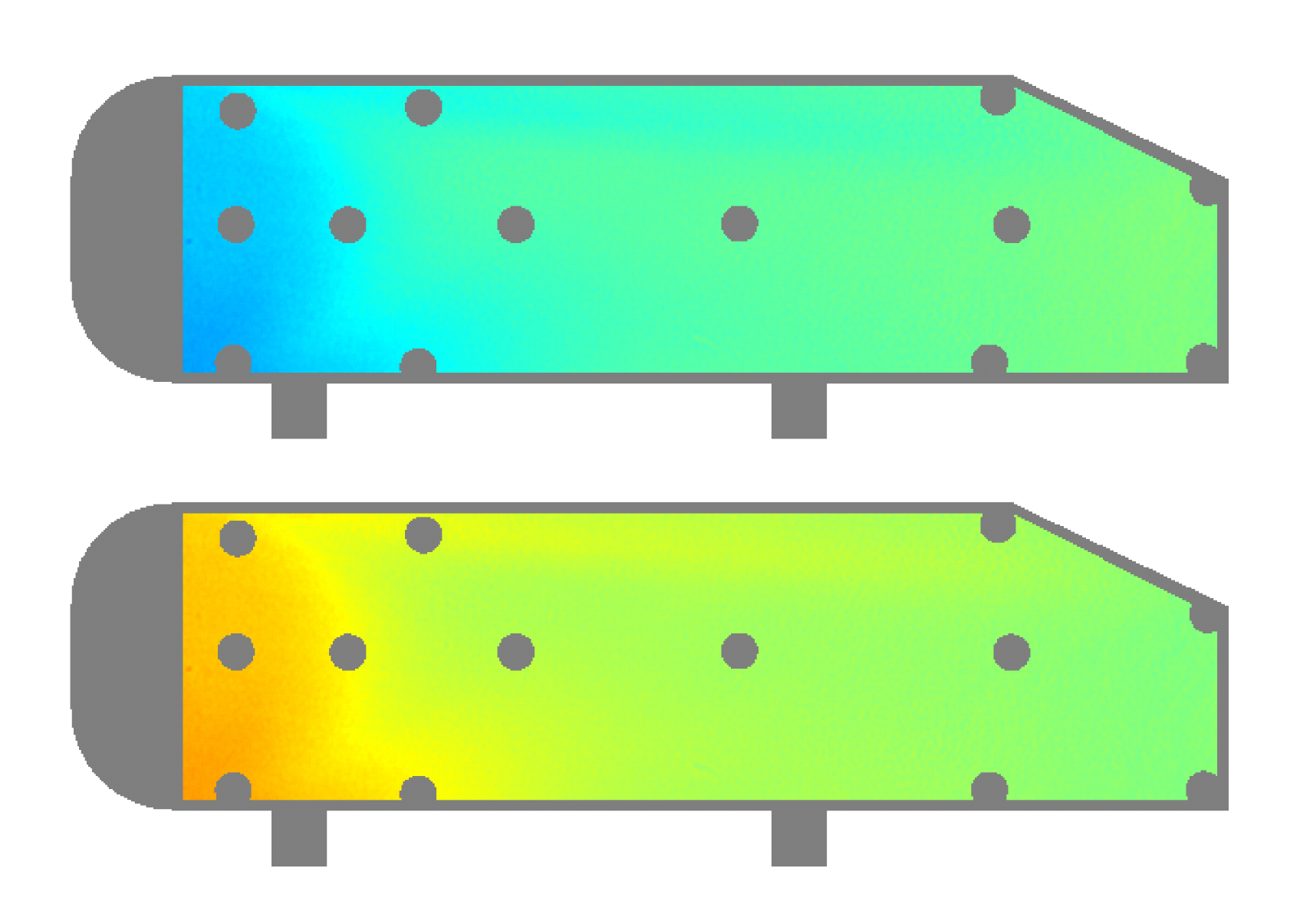}
    \includegraphics[width=0.45\textwidth]{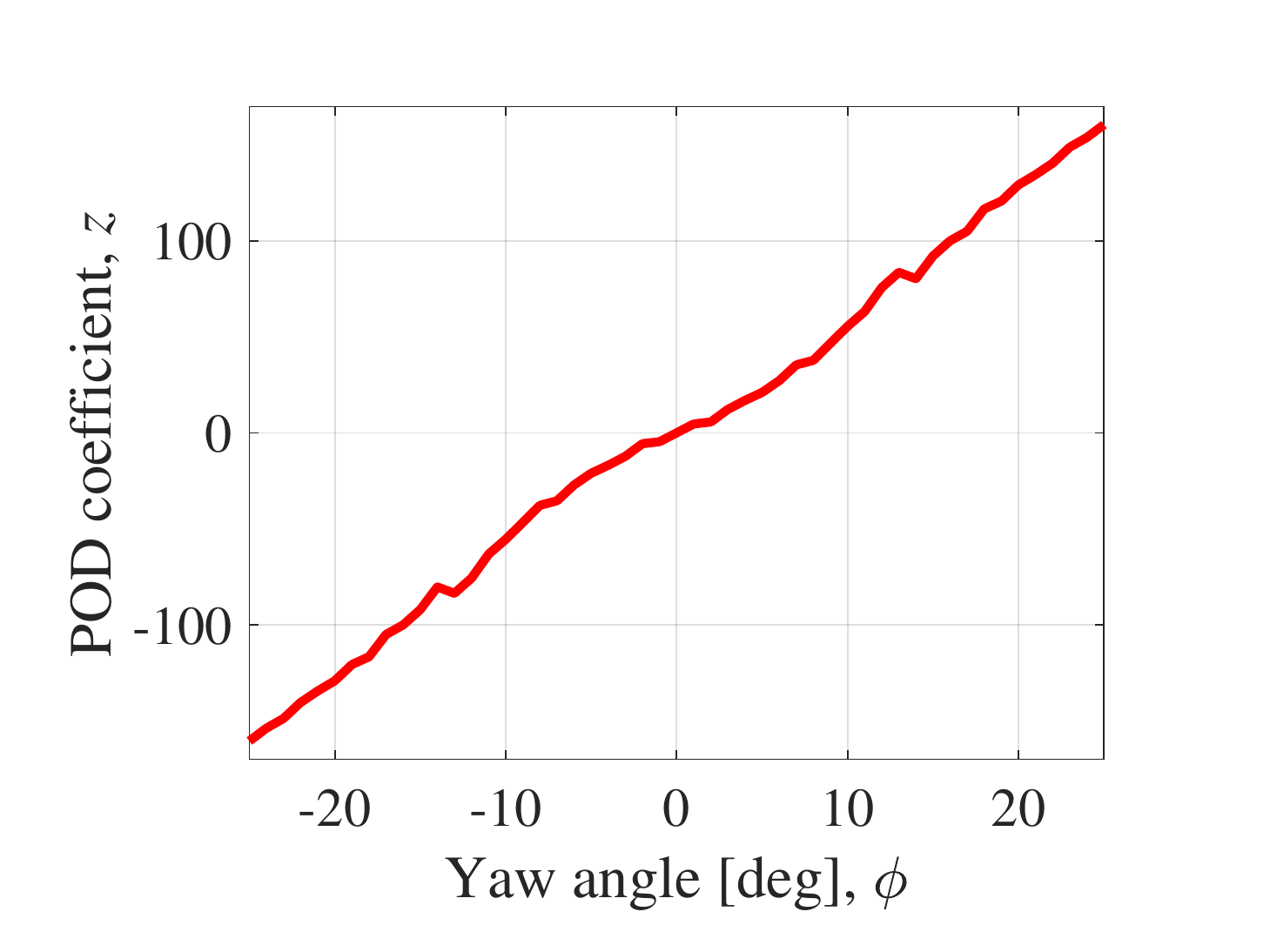}}
    \subfigure[Mode 2]{
    \includegraphics[width=0.45\textwidth]{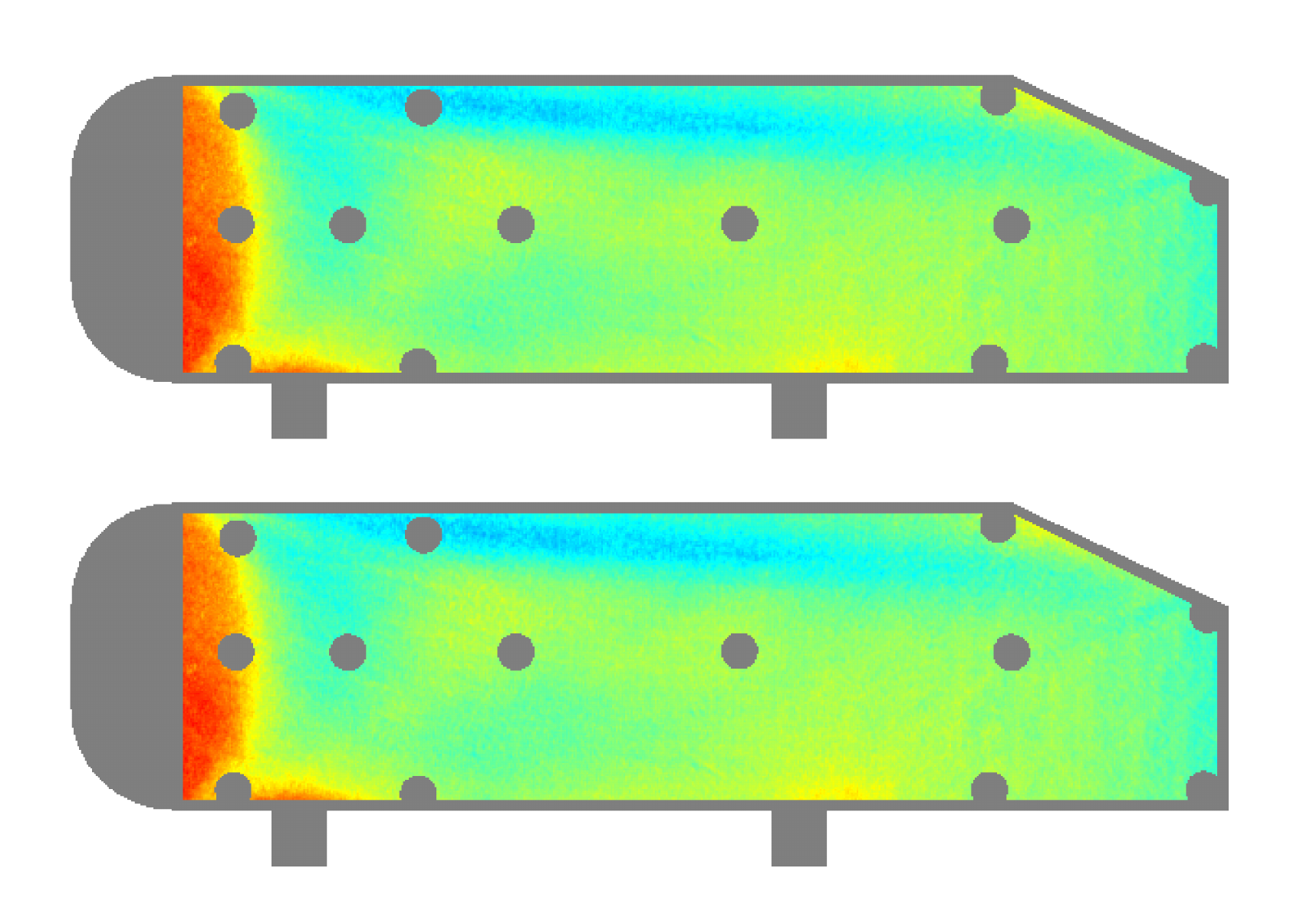}
    \includegraphics[width=0.45\textwidth]{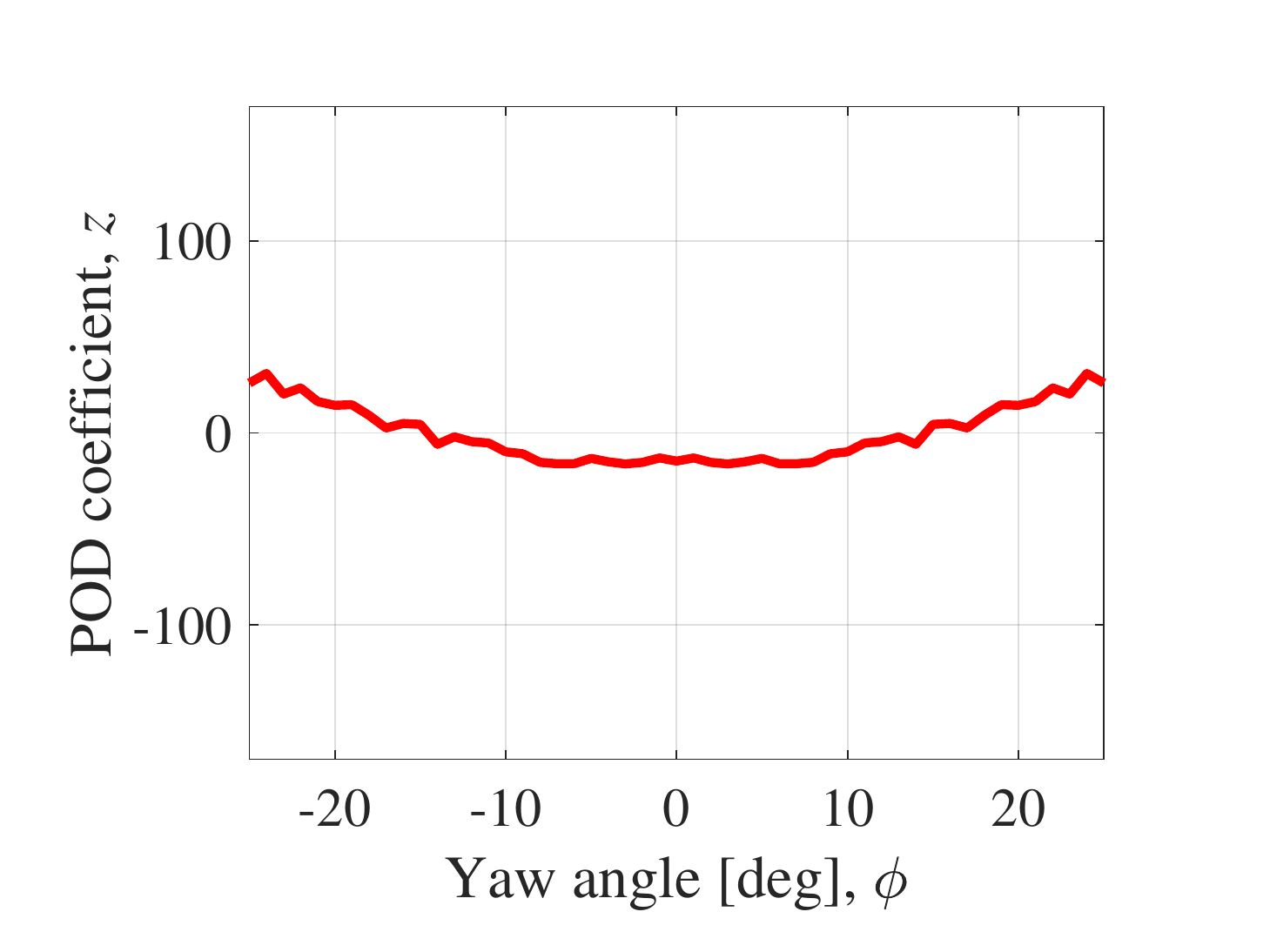}}
\end{figure}

\clearpage

\clearpage
\begin{figure}[!htbp]
    \centering
    \subfigure[The OMP algorithm]{
    \includegraphics[width=0.95\textwidth]{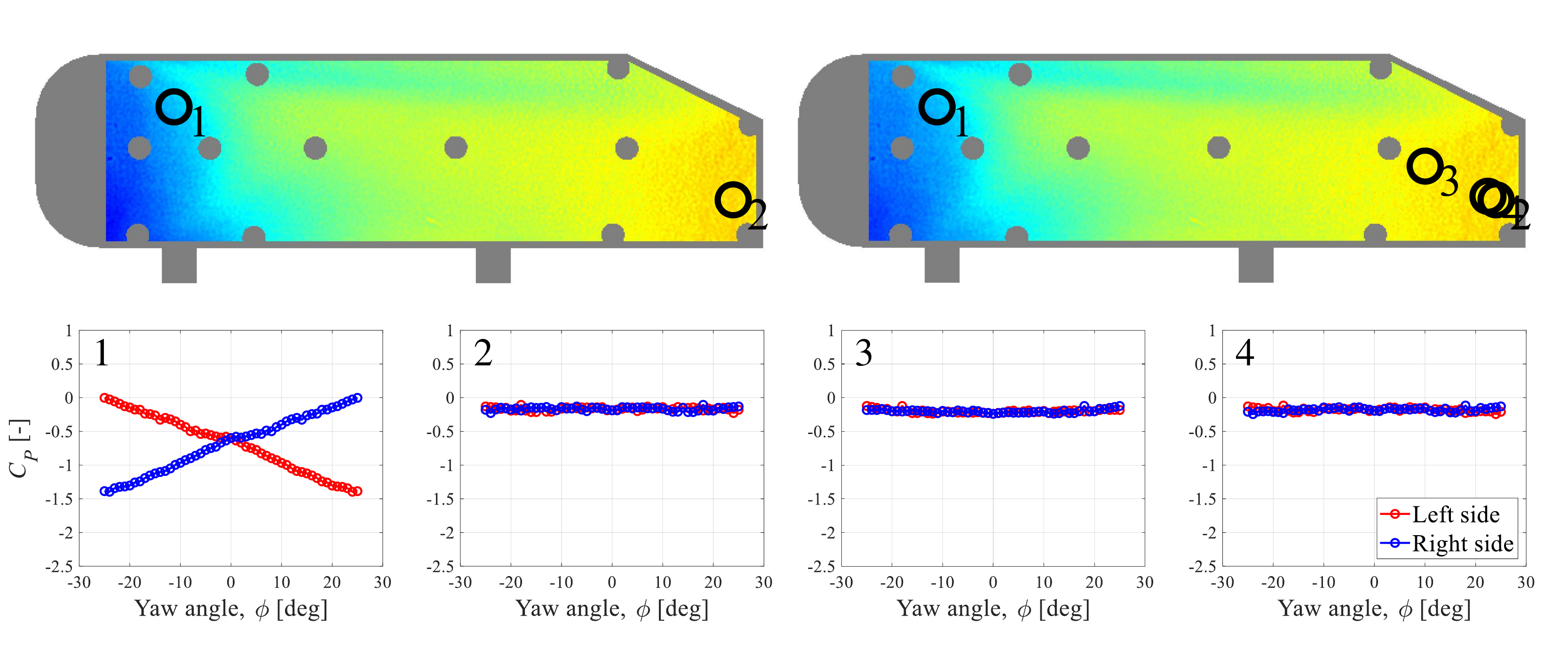}}\\
    \subfigure[The DG-vector algorithm]{
    \includegraphics[width=0.95\textwidth]{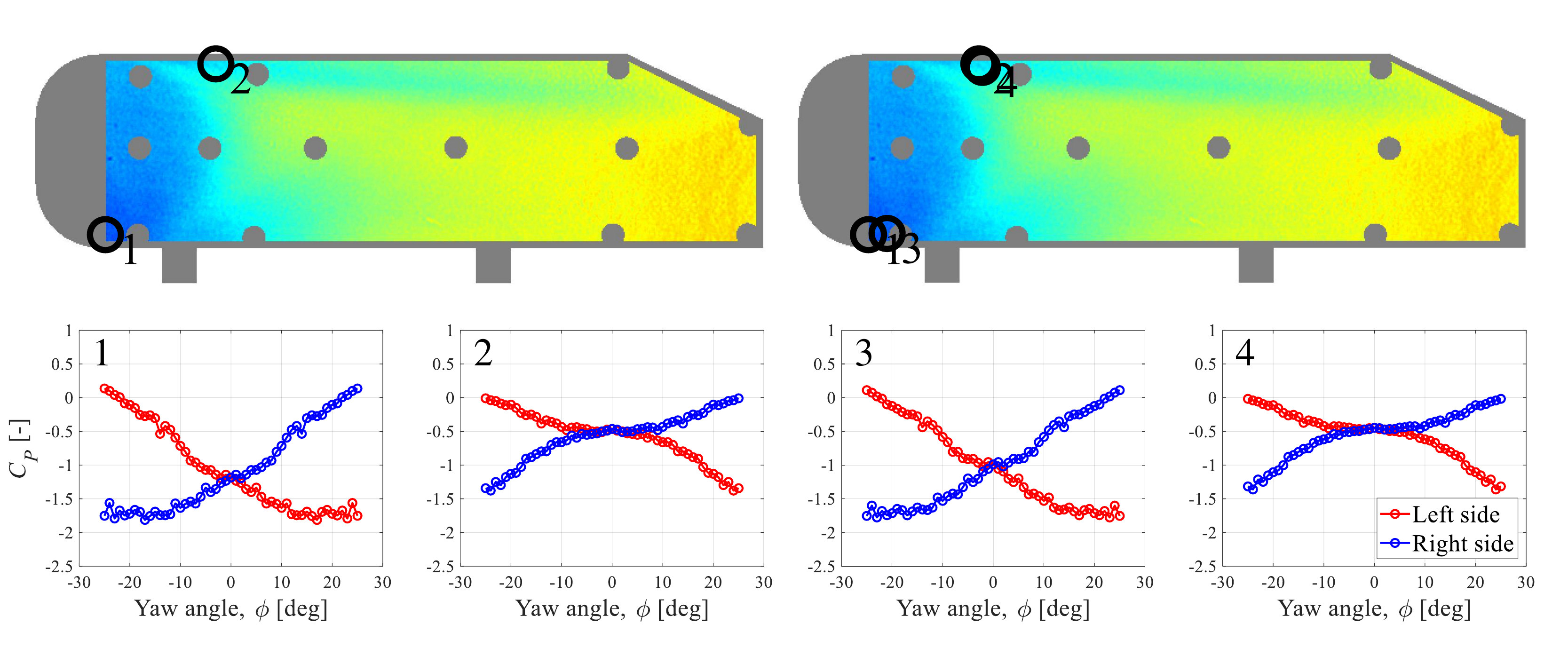}}\\
    \subfigure[The hybrid algorithm]{
    \includegraphics[width=0.95\textwidth]{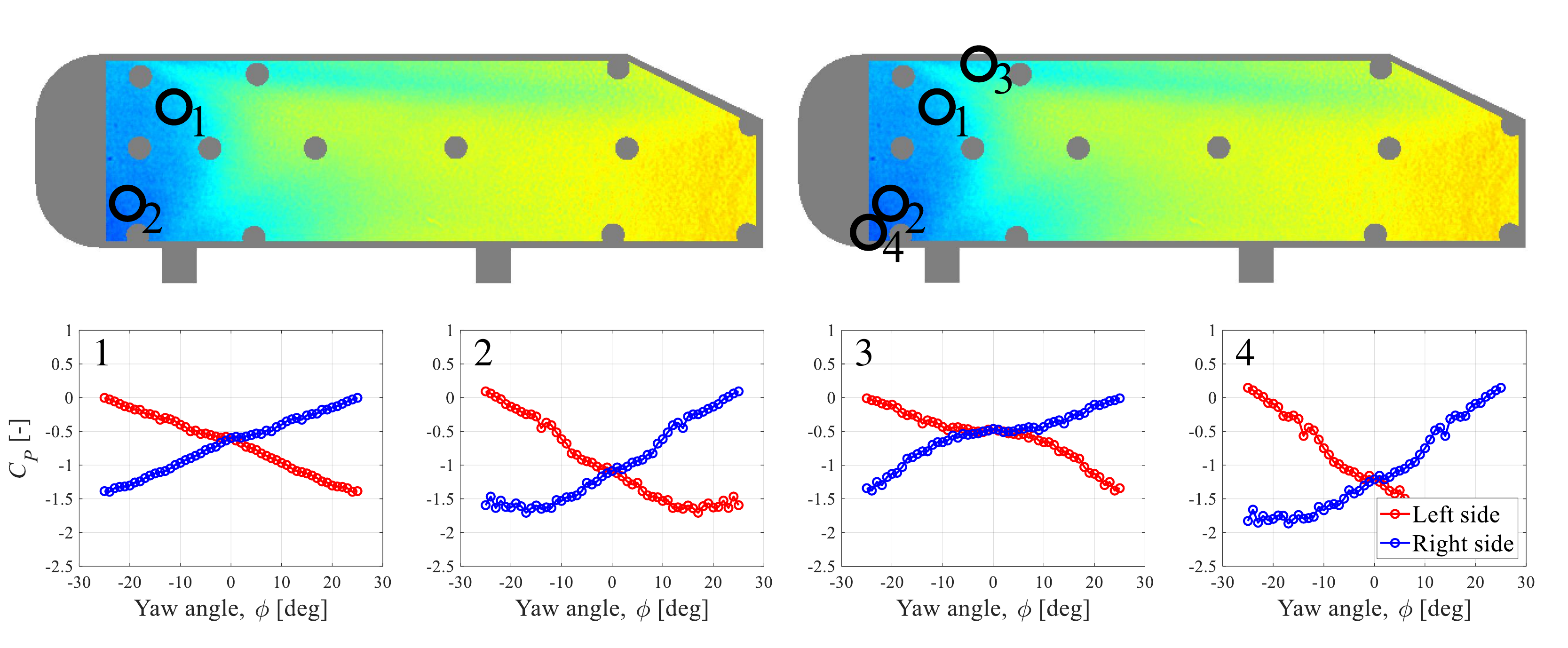}}\\
\end{figure}

\clearpage

\clearpage

\end{document}